\documentclass[a4paper]{article}
\usepackage[ansinew]{inputenc}
\usepackage{times}
\usepackage[T1]{fontenc}
\usepackage{graphicx}
\usepackage{geometry}
\usepackage{amssymb}
\usepackage{amsmath}

\usepackage{rotating}

\usepackage{xcolor}
\colorlet{shadecolor}{gray!25}
\usepackage{subfig}
\usepackage{float}
\usepackage{lscape}

\title{Closed testing procedures for treatment-versus-control comparisons and multiple correlated endpoints}
\author{Ludwig A. Hothorn\\ (retired from) Leibniz University Hannover, Germany, ($ludwig@hothorn.de$) \\
Siegfried Kropf\\ (retired from) Otto-von-Guericke-University Magdeburg, Germany}
				
\date{\today}

\begin{document}
\maketitle
\begin{abstract}
Preferably in two- or three-arm randomized clinical trials, a few (2,3) correlated multiple primary endpoints are considered. In addition to the closed testing principle based on different global tests, two max(maxT) tests are compared with respect to any-pairs, all-pairs and individual power in a simulation study. 
\end{abstract}

\normalsize

\section{The problem}
Multiple endpoints are used in some randomized trials in bio-medicine, whereas the majority are simply independently evaluated univariately, each to level $\alpha$, controlling the comparisonwise error rate (CWER) only. Here, we consider the joint evaluation of a few (e.g., 2, or 3), possibly different-scaled, correlated endpoints. \footnotesize \textit{(Notice, currently manifold published tests for high-dimensional problems, e.g. in genetics, represent a different issue).} 
\normalsize
Two scenarios should be distinguished: i) co-primary endpoints, where the effect evidence is only available if both endpoints contribute with some noncentrality to the desired effect, and ii) primary endpoints, where any endpoint contribute to the effect and a multiple testing procedure allows the interpretation of the marginal hypotheses, i.e. per endpoint (which are the actual hypotheses of interest). This approach is investigated in the context of higher-level global tests for multiple correlated endpoints. Co-primary endpoints (case i))  are analysed be means of an intersection-union test (IUT), where the global claim is proved if each marginal test is under $H_1$, each at level $\alpha$. Primary endpoints (case ii)) are analysed almost differently by means of an union-intersection test (UIT), e.g., by a maxT-t test \cite{Hothorn2020c}. 
Here, the claim is proved if at least one, anyone, marginal test is under $H_1$, each at an adjusted level $\alpha_i$ whereas the entire procedure controls the FWER in a strong sense. Interestingly, the criterion 'at least one, anyone', also includes 'all' marginal tests, which allows a link between case i) and case ii) using the all-pairs power concept).\\
By means of a simulation study several tests are considered for both size and power, whereas first a simple two-sample design with 2 and 3 correlated endpoints is used, followed by a three sample design $[C, T_1, T_2]$ and 2 endpoints. This dimension limitations result from the properties of the closed testing procedure (CTP), which is the focus of the paper. \\
Notice, in biology and genetics, multivariate tests are used to combine the information of many low informative endpoints as stand-alone approaches.

\section{Closed testing procedure based on multivariate global tests}

The closed testing procedure (CTP) \cite{MARCUS1976} and the maxT-test use almost contrary principles for a similar objective. While the maxT-test tests each elementary hypothesis to an adjusted $\alpha$-level (e.g. $\alpha/\xi$)- based on the union-intersection (UIT) principle \textit{('or, ,..., or')}, in the CTP each elementary hypothesis is tested to the full $\alpha$-level, but not alone, rather within an intersection-union (IUT) principle \textit{('and,...,and')}, where all hypotheses in the intersection branch of the closed system tree must also be under $H_1$ (each at $\alpha$-level). The maxT test requires a complex multivariate t-distribution and their correlation matrix, whereas the CTP requires only quantiles of an univariate (F, or t-) distribution (without considering correlation). Central to the CTP is the definition of the elementary hypotheses, which alone are of interest and are then to be interpreted.\\

First, the two-sample design is discussed $H_0: Y_0^1- Y_0^2$. Based on this, all subset intersection hypotheses up to the global hypothesis are constructed, involving  $H_{i0}$. Following the IUT principle, $H_i$ is rejected at level $\alpha$ if and only if $H_i$ itself is rejected and all hypotheses which are include them. Each of these many hypotheses is tested individually with a level $\alpha$-test. This may be any appropriate level $\alpha$-test, making the CPT flexible. Even within a tree, different tests, e.g. F-tests and t-tests, can be used. Each of these branches tests of the tree (determined by the $\xi$ elementary hypotheses) represent an IUT, i.e. $T^{CTP}=min(T_1,...,T_{\xi})$, or $p^{CTP}=max(p_1,...,p_{\xi})$. Comparing of endpoints $Y_{ii'}$ form a complete family of hypotheses \cite{Sonnemann2008} avoiding complex subset intersection hypotheses due to the non-overlapping hypotheses. For the simple bivariate $Y_1,Y_2$ and trivariate $Y_1,Y_2,Y_3$ 2-sample design, the family include the elementary, intersection  and global hypotheses, as shown in Tables 1 and 2

\begin{table}[ht]
\caption{2-sample design: bivariate CTP}
\centering\footnotesize
\begin{tabular}{ll|l|l|l}
No. & Type& $H_0$ & Test  & Tree \\ \hline
1 & Elementary& $H_{0}^1:  \mu_0^1=\mu_1^1$ & $pT_{01}^1$ & T1   \\
2 & Elementary& $H_{0}^2:  \mu_0^2=\mu_1^2$ & $pT_{01}^2$ & T2   \\
3 & Global& $H_{0}^{12}:  \mu_0^{12}=\mu_1^{12}$ & $bivarT_{01}^{12}$ & T1,T2   \\
\hline
\end{tabular}
\end{table}

\begin{table}[ht]
\caption{2-sample design: trivariate CTP}
\centering\footnotesize
\begin{tabular}{ll|l|l|l}
No. & Type& $H_0$ & Test  & Tree \\ \hline
1 & Elementary & $H_{0}^1:  \mu_0^1=\mu_1^1$ & $pT_{01}^1$ & T1   \\
2 & Elementary & $H_{0}^2:  \mu_0^2=\mu_1^2$ & $pT_{01}^2$ & T2   \\
3 & Elementary & $H_{0}^3:  \mu_0^3=\mu_1^3$ & $pT_{01}^2$ & T2   \\
4 & Intersection & $H_{0}^1:H_{0}^2 \Rightarrow H_{0}^{12}:  \mu_0^{12}=\mu_1^{12}$ & $bivarT_{01}^{12}$ & T1,T2   \\
5 & Intersection & $H_{0}^1:H_{0}^3 \Rightarrow H_{0}^{13}:  \mu_0^{13}=\mu_1^{13}$ & $bivarT_{01}^{13}$ & T1,T3   \\
6 & Intersection & $H_{0}^2:H_{0}^3 \Rightarrow H_{0}^{23}:  \mu_0^{23}=\mu_1^{23}$ & $bivarT_{01}^{22}$ & T2,T3   \\
7 & Global& $H_{0}^{123}:  \mu_0^{123}=\mu_1^{123}$ & $trivarT_{01}^{123}$ & T1,T2,T3   \\
    \hline
\end{tabular}
\end{table}

Secondly, considering treatment-versus-control comparisons with multiple endpoints, e.g. the bivariate case, the properties of a complete family are lost and the structure of the CTP tree becomes complex including the partition hypotheses, i.e. the special non-overlapping intersection hypotheses, e.g. for the simple design with $[C, T1, T2], [Y1,Y2]$:

\begin{table}[ht]
\caption{CTP: bivariat C,T1,T2}
\centering\footnotesize
\begin{tabular}{ll|l|l|l}
No. & Type& $H_0$ & Test  & Tree \\ \hline
1 & Elementary& $H_{01}^1$ & $pCT_{01}^1$ & T1   \\
2 & Elementary& $H_{02}^1$ & $pCT_{02}^1$ & T2   \\
3 & Elementary& $H_{01}^2$ & $pCT_{01}^2$ & T3   \\
4 & Elementary& $H_{02}^2$ & $pCT_{02}^2$ & T4   \\ \hline
5 & Intersection of 2 & $H_{01}^1 : H_{02}^1 \Rightarrow H_{012}^1$ & $min(Du^1)$ 								& T1,T2   \\
6 & Intersection of 2 & $H_{01}^{1}: H_{01}^{2} \Rightarrow H_{01}^{12}$ 	& $bivariate T_{01}$ 	&T1,T3    \\
7 & Partition of 2 & $H_{01}^1: H_{02}^2$ 						& $pComb(pCT_{01}^1,pCT_{02}^2)$ 				&T1,T4    \\
8 & Intersection of 2 & $H_{02}^{1}: H_{02}^{2} \Rightarrow H_{02}^{12}$ & $bivariate T_{02}$ 		&T2,T4    \\ 
9 & Intersection of 2 & $H_{01}^{2}: H_{02}^{2} \Rightarrow H_{012}^{2}$ & $min(Du^2)$ 					&T3,T4    \\ 
10 & Partition of 2 & $H_{02}^1:H_{01}^2$ & $pComb(pCT_{02}^1,pCT_{01}^2)$ 										&T2,T3   \\ \hline
11 & Partition of 3 & $H_{01}^1:H_{02}^1: H_{01}^2 \Rightarrow H_{012}^1: H_{01}^2$ & $pComb(min(Du^1),pCT_{01}^2)$  &T1,T2,T3    \\
12 & Partition of 3 & $H_{01}^1:H_{02}^1: H_{02}^2 \Rightarrow H_{012}^1:H_{02}^2 $ & $pComb(min(Du^1),pCT_{02}^2)$  &T1,T2,T4    \\
13 & Partition of 3 & $H_{02}^1:H_{01}^2: H_{02}^2 \Rightarrow H_{02}^1:H_{012}^2$   & $pComb(pCT_{02}^1,min(Du^2)$			&T2,T3,T4   \\
14 & Partition of 3 & $H_{01}^{1}:H_{01}^{2}:H_{02}^2 \Rightarrow H_{01}^1: H_{012}^{2}$ & $pComb(pCT_{01}^1, min(Du^2)$  &T1,T2,T4    \\ \hline
15 & Global & $H_{01}^{1}:H_{02}:H_{02}^1:H_{02}^2 \Rightarrow H_{012}^{1}:H_{012}^2   \Rightarrow H_{012}^{12}$ &   $min(Du^{12})$ &  T1,T2,T3,T4  \\

\end{tabular}
\end{table}

Thus, not only uni-, bi- and tri-variate tests for the 2 and k sample problem are necessary, but also for the specific partition hypotheses for non-overlapping hypotheses, e.g. $H_{01}^{12}: H_{02}^2$, which are tested here using Fisher's product criterion. The already complex system of this simple example shows that CTP is practically limited to only a few comparisons to control (2, 3, or 4) and only a few endpoints 2,3, or 4. In the elementary hypotheses, instead of pairwise t-tests, multiple contrasts for $\mu_i-\mu_0$ are used \cite{Hasler2008, Hothorn2016}, which exploit the entire $df$, a significant power advantage especially for small $n_i$. To avoid directional errors, CTP's were formulated for one-sided hypotheses here. \\

Both CTP and maxT-test become more conservative with increasing number of elementary hypotheses. Controlling FWER, this implies an inherent power loss compared to a single test. In the maxT-test, this is determined by the number $q$ of individual tests in the $\xi$-variate 
t-distribution, with the correlation between the tests having a conservatism-reducing effect. In contrast, the conservativeness of the CTP is due to that of the underlying IUT in the branches of the CTP decision tree, whereas the correlation between the individual tests is not considered in that univariate distribution. Furthermore, the number and type of individual tests in the IUT's depends on the specific structure of the CTP tree. If the number of individual tests is too high  within each IUT (and e.g. 10 are quickly reached in real applications), it becomes hopelessly conservative - a limitation of the CTP in principle to low dimensional cases.

\subsection{Global multivariate tests}
Many multivariate global tests are available, starting with Hotelling's $T^2$.  As here the focus is on one-sided tests with heterogeneous variances between the considered endpoints, we included the parametric and the rank-based version of O'Briens tests  \cite{OBRIEN1984} for multiple endpoints. Additionally, two versions of test statistics based on pairwise distance or similarity measures between sample elements (\cite{Anderson2001}, \cite{Kropf2004}). These tests were originally derived for high-dimensional data and they are based on the idea that sample vectors from different samples should (under the alternative hypothesis) on the average have larger distances or lower similarities than sample vectors form the same sample. Therefore, the average of all pairwise between-group distances (in the present case the squared Euclidean distance and the maximum absolute difference over all variables) is determined and compared with its permutation distribution (399 random permutations). For the present paper, the following adaptations have been made:  (1) To get scale independent tests, all endpoints are standardized by their standard deviation calculated over the pooled sample (which is permutation invariant). (2) One-sided test versions are obtained by including only those endpoints in the calculation of the distance for each pair of sample elements where the treatment element had lower values than the control element and then reject at the lower tail of the permutation distribution. (3) For the test in the three sample design $[C; T1; T2]$, the averages of between-group pairs were calculated separately for $[C; T1]$ and $[C; T2]$ and the maximum of both has been used as test statistics in the permutation test.

\section{MaxT tests for multiple endpoints} \label{Endpoints}
Presumably, the maxT test in connection with multiple endpoints was first proposed  for analysis of multiple binary endpoints as a resampling test \cite{Westfall1989} (albeit as a min-p test; see the comparison between maxT and min-p test \cite{Dudoit2003}). Here, only the parametric maxT test is considered as an asymptotic test version.  The extension to multiple contrast tests with correlated endpoints was introduced as max(maxT) test, where one part of the approximate correlation matrix consists of the calculated correlations between the treatment contrasts and the other part consists of the empirical Pearson correlation between the endpoints \cite{Hasler2011, Hasler2012, Hasler2013, Hasler2018}. Thus, adjusted p-values or simultaneous confidence intervals for both differences and ratios (for control) can be estimated with the CRAN package SimComp.  
An alternative approach for estimating the variance-covariance matrix for  the multivariate normal distribution of underlying  maxT tests represents the multiple marginal models approach \cite{Pipper2012}.   Obtaining the asymptotic joint distributions of parameter estimates from multiple marginal models (glm's up to glmm's) fitted to the same observations is a very helpful and general feature of this approach, available as the function mmm within the CRAN package multcomp \cite{Hothorn2008a}. In the context of multiple endpoints, this approach was used for multiple binary endpoints in both 2- and k-sample design \cite{Klingenberg2013},  for multiple and repeated measures endpoints \cite{Jensen2015, Pallmann2017},   for multiple normal distributed, proportions and different-scaled endpoints \cite{Hothorn2016}(in chapters 2.4.4., 3.6.1. and 5.4), for composite and their marginal endpoints \cite{Ruse2017}, as well for multiple time-to-event endpoints by the Wei-Lachin test \cite{Bebu2018}- probably the first mmm approach.\\

There are some principal properties: i) the power of the max(maxT)) test is equal to that of the Bonferroni test for correlations near to 0 (otherwise for correlations near to 1 of the marginal tests), ii) the more endpoints, the more conservative the procedures, and iii) in the closed test 	are $p_{adjust}=p_{marginal}$ when the p-values in all subset global tests of the IUT branch are smaller.

\section{IUT vs. UIT for testing co-primary endpoints}
Single-step tests for $q$ co-primary endpoints  use commonly the IUT, with the global claim when $p_{adjusted}^{IUT}=max(p_1,...,p_q)< \alpha$. This test has two disadvantages, it is conservative with increasing number of endpoints (because the correlation between the endpoints is not taken into account) and it has only a global statement, i.e. there are no inference statements about the effect of the individual endpoints. Therefore, the alternative considered is the UIT (i.e., exactly the tests for multiple correlated endpoints), but with the all-pairs power concept. The UIT is under $H_1$, if at least one, any single hypothesis is under $H_1$. And this includes the case of interest here, that all single hypotheses are under $H_1$. Using UIT's with the all-pairs power concept provides both the global claim of co-primary endpoints and claims on the individual endpoints by means of adjusted p-values or simultaneous confidence intervals. A simulation study for practically relevant designs must clarify whether the quality disadvantage compared to the IUT is still tolerable.

\section{Simulation study}
Familywise error rate (in weak and strong sense) as well as 4 different power estimates were considered in a simulation study with 5000/10000 independent replications per setting using a two or three sample balanced design ($n_i=5;20$) with normal homoscedastic errors with two (or three) correlated endpoints ($\rho=0.9,0.7,0.5,0$, where $\rho=-0.5$ is without any practical relevance for the unidirectional bivariate tests). Normal distributed, homogeneous errors, complete multivariate data without  missing values, endpoints of different magnitude (so that standardization is needed) and low dimensions (i.e., 2 or 3 treatments, 2,3 endpoints) are used. The non-overlapping hypotheses in the CTP were tested by Fisher product criterion.\\
The aim is to interpret the elementary hypotheses when correlated, normal distributed multiple endpoints are considered.
Therefore, the following tests are compared: i) max(maxT))-test using multivariate t-distribution with Pearson/calculated correlation 
matrix \cite{Hasler2011} ('maxT'), ii) max(maxT))-test using estimated correlation matrix from multiple marginal models \cite{Hothorn2016} ('mmm'), iii) max(maxT)test using Bonferroni-adjustment for multiple endpoints ('Bon'), iv) closed testing using global parametric OBrien-type summary test ('CTPps'), v) closed testing using global rank-transformed OBrien-type summary test ('CTPns'), vi) closed testing using squared Euclidean distance based summary test ('CTPd'), vii) closed testing using maximum absolute component distance based summary test ('CTPD') and viii) IUT ('IUT') when considering all-pairs power only. Furthermore, the marginal tests per endpoint are considered ('Y1,Y2,Y3') as well the individual estimates when considering the individual power (using the extensions 1,2,3).

\subsection{Power definitions}
From the point of view of interpretation, the following 4 power definitions are to be distinguished: i) per-comparison power with control of the comparisonwise error rate CWER(only), ii) individual power, i.e. power per marginal endpoint, iii) any-pairs power (the most common definition in multiple testing) , iv) all-pairs power. Notice, only ii) to iv) control the FWER. For multiple endpoints where the individual per-endpoint inference is of interest, the individual power is considered. This is fundamentally different from any-pairs power (case iii), where one is interested that an endpoint, any endpoint, is under $H_1$. The all-pairs power (case iv) is relevant for co-primary endpoints, where all endpoints must be under $H_1$ for any treatment-versus-control contrast. This is the basic principle of the IUT, but we show below that this is possible for both maxT and CTP under exactly this condition, with additional information available on the marginal hypotheses.\\
Cases (ii) through (iv) are fundamentally different, so a separate section is used for each below (the per-comparisonwise power is also shown in each case for purpose).

\subsection{Two sample design with 2 correlated endpoints}
\subsubsection{Two sample design with 2 endpoints: any-pairs power} \label{ANY}
Table \ref{tab:any} shows FWER-estimates (under global $H_0$), the strong control of FWER (under partial $H_{01}$) 
and the any-pairs power for several noncentralities of both endpoints and several $\rho$. Per definition, all multiple tests (maxT-tests, CTP) control FWER in both weak and strong sense, with the tendency of conservatism for Bonferroni towards high correlation, whereas CTP towards low correlation (an IUT property, see chapter \ref{ALL}). The power behavior is quite deficient for different non-centralities $\delta$ and various $\rho$ in different tests. For too high $\delta$, all powers approach the value of 1, so this unrealistic scenario is ignored. \\

Not surprising, no uniform most powerful test (umpt) exists for these different configurations. In some circumstances,  the any-pairs power can be higher than any of the comparisonwise power (based on CWER control only)- particularly when the both not-too-large noncentralities of both endpoints are similar, but only in the direction of lower correlation, i.e. $\rho<0.7$. Exactly for this configuration, the any-pairs power of the CTP (particularly maximum component distance based test is little higher than that of the maxT tests, vice versa.  This exactly may be practically relevant: both already correlated endpoints contribute to some extend to the effect evidence. However, this power advantage over the Bonferroni test is not too large (not surprising for theses low $\rho$). The any-pairs power of the CTP is seriously reduced when the noncentrality of the other endpoint is small or even trends to zero (due to the IUT property). 

\begin{table}[H]
\centering\small
\caption{Bivariate two-sample design: any-pairs power}
\label{tab:any}
\scalebox{0.65}{
\begin{tabular}{rrrrrrrrr|rr|rr|rr|r|rr}
  \hline
 & n & $\mu_{11}$ & $\mu_{12}$ & $\mu_{21}$ & $\mu_{22}$ & sd1 & sd2 & rho & maxT & mmm & CTPns & CTPd & CTPD & CTPps & Bon & Y1 & Y2 \\ 
  \hline
$H_0$ & 20 & 1 & 1 & 10 & 10 & 5.0 & 10.0 & 0.9 & 0.05 & 0.05 & 0.04 & 0.04 & 0.04 & 0.05 & 0.03 & 0.05 & 0.05 \\ 
 & 20 & 1 & 1 & 10 & 10 & 5.0 & 10.0 & 0.7 & 0.05 & 0.05 & 0.05 & 0.05 & 0.05 & 0.05 & 0.04 & 0.05 & 0.05 \\ 
 & 20 & 1 & 1 & 10 & 10 & 5.0 & 10.0 & 0.0 & 0.05 & 0.05 & 0.03 & 0.03 & 0.03 & 0.03 & 0.05 & 0.05 & 0.05 \\ 
 & 20 & 1 & 1 & 10 & 10 & 5.0 & 10.0 & -0.5 & 0.05 & 0.05 & 0.03 & 0.02 & 0.02 & 0.03 & 0.05 & 0.05 & 0.05 \\ 
\hline\hline
$H_{0,1}$ & 20 & 1 & 1 & 10 & 20 & 5.0 & 10.0 & 0.9 & 0.91 & 0.91 & 0.45 & 0.32 & 0.24 & 0.45 & 0.88 & 0.05 & 0.93 \\ 
 & 20 & 1 & 1 & 10 & 18 & 5.0 & 10.0 & 0.9 & 0.76 & 0.76 & 0.33 & 0.25 & 0.20 & 0.33 & 0.70 & 0.05 & 0.81 \\ 
 & 20 & 1 & 1 & 10 & 17 & 5.0 & 10.0 & 0.9 & 0.64 & 0.64 & 0.28 & 0.22 & 0.19 & 0.29 & 0.57 & 0.05 & 0.70 \\ 
 & 20 & 1 & 1 & 10 & 16 & 5.0 & 10.0 & 0.9 & 0.53 & 0.53 & 0.23 & 0.19 & 0.16 & 0.23 & 0.46 & 0.05 & 0.58 \\ 
\hline
 & 20 & 1 & 1 & 10 & 20 & 5.0 & 10.0 & 0.7 & 0.88 & 0.88 & 0.47 & 0.33 & 0.30 & 0.47 & 0.86 & 0.05 & 0.92 \\ 
 & 20 & 1 & 1 & 10 & 18 & 5.0 & 10.0 & 0.7 & 0.72 & 0.72 & 0.36 & 0.27 & 0.26 & 0.36 & 0.69 & 0.05 & 0.79 \\ 
 & 20 & 1 & 1 & 10 & 17 & 5.0 & 10.0 & 0.7 & 0.61 & 0.62 & 0.29 & 0.24 & 0.23 & 0.30 & 0.58 & 0.05 & 0.70 \\ 
 & 20 & 1 & 1 & 10 & 16 & 5.0 & 10.0 & 0.7 & 0.49 & 0.49 & 0.25 & 0.20 & 0.20 & 0.24 & 0.46 & 0.05 & 0.59 \\ 
\hline
 & 20 & 1 & 1 & 10 & 20 & 5.0 & 10.0 & 0.5 & 0.87 & 0.87 & 0.52 & 0.37 & 0.36 & 0.52 & 0.86 & 0.05 & 0.92 \\ 
 & 20 & 1 & 1 & 10 & 18 & 5.0 & 10.0 & 0.5 & 0.71 & 0.71 & 0.39 & 0.30 & 0.30 & 0.39 & 0.69 & 0.05 & 0.79 \\ 
 & 20 & 1 & 1 & 10 & 17 & 5.0 & 10.0 & 0.5 & 0.60 & 0.60 & 0.33 & 0.27 & 0.26 & 0.33 & 0.58 & 0.05 & 0.70 \\ 
 & 20 & 1 & 1 & 10 & 16 & 5.0 & 10.0 & 0.5 & 0.47 & 0.48 & 0.26 & 0.22 & 0.22 & 0.27 & 0.46 & 0.05 & 0.58 \\ 
\hline
 & 20 & 1 & 1 & 10 & 20 & 5.0 & 10.0 & 0.0 & 0.87 & 0.87 & 0.66 & 0.48 & 0.52 & 0.66 & 0.87 & 0.05 & 0.93 \\ 
 & 20 & 1 & 1 & 10 & 18 & 5.0 & 10.0 & 0.0 & 0.70 & 0.70 & 0.49 & 0.38 & 0.41 & 0.50 & 0.70 & 0.05 & 0.80 \\ 
 & 20 & 1 & 1 & 10 & 17 & 5.0 & 10.0 & 0.0 & 0.57 & 0.58 & 0.39 & 0.31 & 0.33 & 0.40 & 0.57 & 0.05 & 0.69 \\ 
 & 20 & 1 & 1 & 10 & 16 & 5.0 & 10.0 & 0.0 & 0.48 & 0.48 & 0.33 & 0.26 & 0.28 & 0.34 & 0.48 & 0.06 & 0.60 \\ 
\hline\hline
 $H_{1,1}$& 20 & 1 & 5 & 10 & 20 & 5.0 & 10.0 & 0.9 & 0.90 & 0.90 & 0.87 & 0.86 & 0.87 & 0.89 & 0.87 & 0.79 & 0.92 \\ 
 & 20 & 1 & 5 & 10 & 18 & 5.0 & 10.0 & 0.9 & 0.81 & 0.81 & 0.79 & 0.78 & 0.80 & 0.81 & 0.76 & 0.80 & 0.80 \\ 
 & 20 & 1 & 5 & 10 & 17 & 5.0 & 10.0 & 0.9 & 0.78 & 0.78 & 0.76 & 0.74 & 0.75 & 0.78 & 0.72 & 0.81 & 0.70 \\ 
 & 20 & 1 & 5 & 10 & 16 & 5.0 & 10.0 & 0.9 & 0.76 & 0.76 & 0.69 & 0.69 & 0.69 & 0.71 & 0.70 & 0.80 & 0.59 \\ 
\hline
 & 20 & 1 & 5 & 10 & 20 & 5.0 & 10.0 & 0.7 & 0.91 & 0.92 & 0.91 & 0.89 & 0.90 & 0.92 & 0.89 & 0.80 & 0.93 \\ 
 & 20 & 1 & 5 & 10 & 18 & 5.0 & 10.0 & 0.7 & 0.83 & 0.83 & 0.83 & 0.82 & 0.83 & 0.84 & 0.80 & 0.79 & 0.80 \\ 
 & 20 & 1 & 5 & 10 & 17 & 5.0 & 10.0 & 0.7 & 0.80 & 0.80 & 0.79 & 0.78 & 0.79 & 0.80 & 0.76 & 0.80 & 0.70 \\ 
 & 20 & 1 & 5 & 10 & 16 & 5.0 & 10.0 & 0.7 & 0.76 & 0.76 & 0.73 & 0.72 & 0.73 & 0.75 & 0.73 & 0.80 & 0.59 \\ 
\hline
 & 20 & 1 & 5 & 10 & 20 & 5.0 & 10.0 & 0.5 & 0.93 & 0.93 & 0.94 & 0.93 & 0.94 & 0.94 & 0.93 & 0.81 & 0.94 \\ 
 & 20 & 1 & 5 & 10 & 18 & 5.0 & 10.0 & 0.5 & 0.86 & 0.86 & 0.87 & 0.86 & 0.87 & 0.88 & 0.84 & 0.80 & 0.80 \\ 
 & 20 & 1 & 5 & 10 & 17 & 5.0 & 10.0 & 0.5 & 0.81 & 0.81 & 0.82 & 0.81 & 0.82 & 0.83 & 0.79 & 0.80 & 0.69 \\ 
 & 20 & 1 & 5 & 10 & 16 & 5.0 & 10.0 & 0.5 & 0.78 & 0.79 & 0.78 & 0.77 & 0.78 & 0.80 & 0.77 & 0.80 & 0.59 \\ 
\hline
 & 20 & 1 & 5 & 10 & 20 & 5.0 & 10.0 & 0.0 & 0.96 & 0.96 & 0.98 & 0.97 & 0.98 & 0.98 & 0.96 & 0.79 & 0.93 \\ 
 & 20 & 1 & 5 & 10 & 18 & 5.0 & 10.0 & 0.0 & 0.91 & 0.91 & 0.94 & 0.94 & 0.94 & 0.95 & 0.91 & 0.81 & 0.80 \\ 
 & 20 & 1 & 5 & 10 & 17 & 5.0 & 10.0 & 0.0 & 0.87 & 0.87 & 0.92 & 0.91 & 0.92 & 0.92 & 0.87 & 0.79 & 0.71 \\ 
 & 20 & 1 & 5 & 10 & 16 & 5.0 & 10.0 & 0.0 & 0.82 & 0.83 & 0.88 & 0.86 & 0.87 & 0.88 & 0.82 & 0.80 & 0.58 \\ 
\hline
 & 20 & 1 & 5 & 10 & 20 & 5.0 & 10.0 & -0.5 & 0.99 & 0.99 & 1.00 & 1.00 & 1.00 & 1.00 & 0.99 & 0.80 & 0.93 \\ 
 & 20 & 1 & 5 & 10 & 18 & 5.0 & 10.0 & -0.5 & 0.96 & 0.96 & 0.99 & 0.99 & 0.99 & 0.99 & 0.96 & 0.80 & 0.80 \\ 
 & 20 & 1 & 5 & 10 & 17 & 5.0 & 10.0 & -0.5 & 0.94 & 0.94 & 0.98 & 0.98 & 0.98 & 0.98 & 0.94 & 0.81 & 0.70 \\ 
 & 20 & 1 & 5 & 10 & 16 & 5.0 & 10.0 & -0.5 & 0.90 & 0.90 & 0.96 & 0.95 & 0.96 & 0.96 & 0.90 & 0.80 & 0.59 \\ 
\hline\hline
$H_{\delta,\delta}$ & 20 & 1 & 5 & 10 & 20 & 5.0 & 10.0 & 0.9 & 0.91 & 0.91 & 0.87 & 0.86 & 0.87 & 0.89 & 0.88 & 0.80 & 0.93 \\ 
 & 20 & 1 & 4 & 10 & 18 & 5.0 & 10.0 & 0.9 & 0.76 & 0.76 & 0.70 & 0.69 & 0.70 & 0.72 & 0.71 & 0.59 & 0.80 \\ 
 & 20 & 1 & 3 & 10 & 17 & 5.0 & 10.0 & 0.9 & 0.65 & 0.65 & 0.54 & 0.51 & 0.51 & 0.55 & 0.59 & 0.36 & 0.71 \\ 
 & 20 & 1 & 2 & 10 & 16 & 5.0 & 10.0 & 0.9 & 0.53 & 0.53 & 0.34 & 0.31 & 0.30 & 0.35 & 0.46 & 0.15 & 0.59 \\ 
\hline
 & 20 & 1 & 5 & 10 & 20 & 5.0 & 10.0 & 0.7 & 0.92 & 0.92 & 0.90 & 0.90 & 0.90 & 0.92 & 0.90 & 0.80 & 0.93 \\ 
 & 20 & 1 & 4 & 10 & 18 & 5.0 & 10.0 & 0.7 & 0.77 & 0.77 & 0.74 & 0.73 & 0.74 & 0.76 & 0.74 & 0.59 & 0.81 \\ 
 & 20 & 1 & 3 & 10 & 17 & 5.0 & 10.0 & 0.7 & 0.62 & 0.63 & 0.55 & 0.53 & 0.53 & 0.57 & 0.59 & 0.34 & 0.69 \\ 
 & 20 & 1 & 2 & 10 & 16 & 5.0 & 10.0 & 0.7 & 0.51 & 0.51 & 0.37 & 0.33 & 0.33 & 0.38 & 0.47 & 0.16 & 0.60 \\ 
\hline
 & 20 & 1 & 5 & 10 & 20 & 5.0 & 10.0 & 0.0 & 0.96 & 0.96 & 0.98 & 0.97 & 0.98 & 0.98 & 0.96 & 0.80 & 0.92 \\ 
 & 20 & 1 & 4 & 10 & 18 & 5.0 & 10.0 & 0.0 & 0.82 & 0.83 & 0.88 & 0.86 & 0.87 & 0.88 & 0.82 & 0.58 & 0.79 \\ 
 & 20 & 1 & 3 & 10 & 17 & 5.0 & 10.0 & 0.0 & 0.68 & 0.68 & 0.71 & 0.68 & 0.69 & 0.73 & 0.68 & 0.34 & 0.71 \\ 
 & 20 & 1 & 2 & 10 & 16 & 5.0 & 10.0 & 0.0 & 0.51 & 0.51 & 0.48 & 0.43 & 0.45 & 0.49 & 0.51 & 0.14 & 0.59 \\ 
\hline
 & 20 & 1 & 5 & 10 & 20 & 5.0 & 10.0 & -0.5 & 0.99 & 0.99 & 1.00 & 1.00 & 1.00 & 1.00 & 0.99 & 0.80 & 0.93 \\ 
 & 20 & 1 & 4 & 10 & 18 & 5.0 & 10.0 & -0.5 & 0.90 & 0.91 & 0.96 & 0.95 & 0.96 & 0.96 & 0.91 & 0.58 & 0.79 \\ 
 & 20 & 1 & 3 & 10 & 17 & 5.0 & 10.0 & -0.5 & 0.74 & 0.74 & 0.85 & 0.82 & 0.84 & 0.86 & 0.74 & 0.35 & 0.71 \\ 
 & 20 & 1 & 2 & 10 & 16 & 5.0 & 10.0 & -0.5 & 0.52 & 0.53 & 0.60 & 0.55 & 0.58 & 0.62 & 0.53 & 0.15 & 0.58 \\ 
   \hline
\end{tabular}
}
\end{table}

\subsubsection{Two sample design with 2 endpoints: individual power} \label{IND}
Table \ref{tab:ind} reveals none of the individual power of maxT or CTP  is the same or even larger than the marginal power when controlling CWER instead of FWER only - a basic property of simultaneous inference. Both maxT and CTP pay a multiplicity price for controlling FWER in the multiple endpoint context as well. The use of powerful global multivariate tests does not change this basic property.

\begin{table}[H]
\centering\footnotesize
\caption{Bivariate two-sample design: individual power}
\label{tab:ind}
\scalebox{0.5}{
\begin{tabular}{rrrrrrrrr|rr|rr|rr|rr|rr|rr|rr|rr}
  \hline
 & n & ma1 & ma2 & mb1 & mb2 & sa & sb & rho & maxT1 & maxT2 & mmm1 & mmm2 & CTPns1 & CTPns2 & Bon1 & Bon2 & CTPd1 & CTPd2 & CTPD1 & CTPD2 & CTPsp1 & CTPsp1 & Y1 & Y2 \\
  \hline
$H_0$ & 20 & 1 & 1 & 10 & 10 & 5.0 & 10.0 & 0.9 & 0.03 & 0.04 & 0.03 & 0.04 & 0.04 & 0.04 & 0.02 & 0.03 & 0.04 & 0.04 & 0.04 & 0.04 & 0.04 & 0.04 & 0.05 & 0.05 \\
 & 20 & 1 & 1 & 10 & 10 & 5.0 & 10.0 & 0.7 & 0.03 & 0.03 & 0.03 & 0.03 & 0.03 & 0.03 & 0.03 & 0.02 & 0.03 & 0.03 & 0.03 & 0.03 & 0.03 & 0.03 & 0.05 & 0.05 \\
 & 20 & 1 & 1 & 10 & 10 & 5.0 & 10.0 & 0.0 & 0.02 & 0.02 & 0.02 & 0.03 & 0.02 & 0.02 & 0.02 & 0.02 & 0.02 & 0.02 & 0.02 & 0.02 & 0.02 & 0.02 & 0.05 & 0.05 \\
 & 20 & 1 & 1 & 10 & 10 & 5.0 & 10.0 & -0.5 & 0.02 & 0.03 & 0.02 & 0.03 & 0.01 & 0.01 & 0.02 & 0.03 & 0.01 & 0.01 & 0.01 & 0.01 & 0.01 & 0.01 & 0.05 & 0.05 \\
\hline\hline
$H_{0,1}$ & 20 & 1 & 1 & 10 & 20 & 5.0 & 10.0 & 0.9 & 0.03 & 0.90 & 0.03 & 0.90 & 0.04 & 0.44 & 0.02 & 0.87 & 0.04 & 0.30 & 0.04 & 0.23 & 0.04 & 0.43 & 0.04 & 0.93 \\
 & 20 & 1 & 1 & 10 & 18 & 5.0 & 10.0 & 0.9 & 0.04 & 0.76 & 0.04 & 0.76 & 0.05 & 0.32 & 0.03 & 0.70 & 0.05 & 0.25 & 0.05 & 0.20 & 0.05 & 0.33 & 0.05 & 0.80 \\
 & 20 & 1 & 1 & 10 & 17 & 5.0 & 10.0 & 0.9 & 0.03 & 0.63 & 0.03 & 0.63 & 0.05 & 0.27 & 0.02 & 0.57 & 0.05 & 0.21 & 0.05 & 0.18 & 0.05 & 0.27 & 0.05 & 0.69 \\
 & 20 & 1 & 1 & 10 & 16 & 5.0 & 10.0 & 0.9 & 0.03 & 0.52 & 0.03 & 0.52 & 0.05 & 0.23 & 0.02 & 0.45 & 0.05 & 0.19 & 0.05 & 0.16 & 0.05 & 0.22 & 0.05 & 0.58 \\
\hline
& 20 & 1 & 1 & 10 & 20 & 5.0 & 10.0 & 0.7 & 0.03 & 0.88 & 0.03 & 0.89 & 0.05 & 0.48 & 0.03 & 0.87 & 0.05 & 0.33 & 0.05 & 0.30 & 0.05 & 0.47 & 0.05 & 0.92 \\
 & 20 & 1 & 1 & 10 & 18 & 5.0 & 10.0 & 0.7 & 0.03 & 0.72 & 0.03 & 0.72 & 0.05 & 0.35 & 0.02 & 0.69 & 0.05 & 0.27 & 0.05 & 0.25 & 0.05 & 0.35 & 0.05 & 0.80 \\
 & 20 & 1 & 1 & 10 & 17 & 5.0 & 10.0 & 0.7 & 0.03 & 0.62 & 0.03 & 0.62 & 0.05 & 0.31 & 0.02 & 0.59 & 0.05 & 0.24 & 0.05 & 0.23 & 0.05 & 0.30 & 0.05 & 0.71 \\
 & 20 & 1 & 1 & 10 & 16 & 5.0 & 10.0 & 0.7 & 0.03 & 0.48 & 0.03 & 0.48 & 0.05 & 0.24 & 0.03 & 0.44 & 0.05 & 0.20 & 0.05 & 0.19 & 0.05 & 0.24 & 0.05 & 0.58 \\
\hline
 & 20 & 1 & 1 & 10 & 20 & 5.0 & 10.0 & 0.5 & 0.03 & 0.88 & 0.03 & 0.88 & 0.05 & 0.52 & 0.03 & 0.86 & 0.05 & 0.37 & 0.05 & 0.37 & 0.05 & 0.52 & 0.05 & 0.93 \\
 & 20 & 1 & 1 & 10 & 18 & 5.0 & 10.0 & 0.5 & 0.02 & 0.71 & 0.02 & 0.72 & 0.05 & 0.40 & 0.02 & 0.70 & 0.05 & 0.30 & 0.05 & 0.30 & 0.05 & 0.40 & 0.05 & 0.80 \\
 & 20 & 1 & 1 & 10 & 17 & 5.0 & 10.0 & 0.5 & 0.02 & 0.60 & 0.02 & 0.60 & 0.05 & 0.32 & 0.02 & 0.57 & 0.05 & 0.25 & 0.05 & 0.25 & 0.05 & 0.32 & 0.05 & 0.70 \\
 & 20 & 1 & 1 & 10 & 16 & 5.0 & 10.0 & 0.5 & 0.03 & 0.46 & 0.03 & 0.46 & 0.05 & 0.26 & 0.02 & 0.45 & 0.05 & 0.22 & 0.05 & 0.22 & 0.05 & 0.27 & 0.05 & 0.57 \\
\hline
 & 20 & 1 & 1 & 10 & 20 & 5.0 & 10.0 & 0.0 & 0.02 & 0.87 & 0.03 & 0.87 & 0.05 & 0.65 & 0.02 & 0.87 & 0.05 & 0.48 & 0.05 & 0.52 & 0.05 & 0.65 & 0.05 & 0.93 \\
 & 20 & 1 & 1 & 10 & 18 & 5.0 & 10.0 & 0.0 & 0.03 & 0.70 & 0.03 & 0.70 & 0.05 & 0.49 & 0.03 & 0.70 & 0.05 & 0.37 & 0.05 & 0.40 & 0.05 & 0.49 & 0.05 & 0.80 \\
 & 20 & 1 & 1 & 10 & 17 & 5.0 & 10.0 & 0.0 & 0.02 & 0.58 & 0.02 & 0.59 & 0.05 & 0.39 & 0.02 & 0.58 & 0.05 & 0.31 & 0.05 & 0.33 & 0.05 & 0.40 & 0.05 & 0.70 \\
 & 20 & 1 & 1 & 10 & 16 & 5.0 & 10.0 & 0.0 & 0.03 & 0.47 & 0.03 & 0.47 & 0.05 & 0.32 & 0.03 & 0.47 & 0.05 & 0.26 & 0.05 & 0.27 & 0.05 & 0.33 & 0.05 & 0.59 \\
\hline\hline

$H_{1,1}$ & 20 & 1 & 5 & 10 & 20 & 5.0 & 10.0 & 0.9 & 0.76 & 0.91 & 0.76 & 0.91 & 0.80 & 0.88 & 0.69 & 0.87 & 0.79 & 0.86 & 0.80 & 0.87 & 0.80 & 0.89 & 0.80 & 0.93 \\
 & 20 & 1 & 5 & 10 & 18 & 5.0 & 10.0 & 0.9 & 0.75 & 0.75 & 0.75 & 0.75 & 0.76 & 0.76 & 0.69 & 0.69 & 0.76 & 0.76 & 0.77 & 0.77 & 0.78 & 0.78 & 0.80 & 0.79 \\
 & 20 & 1 & 5 & 10 & 17 & 5.0 & 10.0 & 0.9 & 0.75 & 0.64 & 0.76 & 0.65 & 0.74 & 0.68 & 0.69 & 0.58 & 0.72 & 0.67 & 0.74 & 0.69 & 0.76 & 0.70 & 0.80 & 0.70 \\
 & 20 & 1 & 5 & 10 & 16 & 5.0 & 10.0 & 0.9 & 0.75 & 0.53 & 0.75 & 0.53 & 0.70 & 0.58 & 0.69 & 0.46 & 0.68 & 0.57 & 0.69 & 0.58 & 0.71 & 0.58 & 0.80 & 0.58 \\
\hline
 & 20 & 1 & 5 & 10 & 20 & 5.0 & 10.0 & 0.7 & 0.72 & 0.89 & 0.72 & 0.89 & 0.80 & 0.89 & 0.69 & 0.87 & 0.79 & 0.88 & 0.80 & 0.88 & 0.80 & 0.90 & 0.80 & 0.93 \\
 & 20 & 1 & 5 & 10 & 18 & 5.0 & 10.0 & 0.7 & 0.72 & 0.71 & 0.73 & 0.72 & 0.77 & 0.76 & 0.69 & 0.68 & 0.76 & 0.76 & 0.77 & 0.77 & 0.78 & 0.77 & 0.80 & 0.80 \\
 & 20 & 1 & 5 & 10 & 17 & 5.0 & 10.0 & 0.7 & 0.73 & 0.61 & 0.73 & 0.61 & 0.74 & 0.68 & 0.70 & 0.57 & 0.73 & 0.67 & 0.75 & 0.68 & 0.76 & 0.68 & 0.80 & 0.70 \\
 & 20 & 1 & 5 & 10 & 16 & 5.0 & 10.0 & 0.7 & 0.73 & 0.50 & 0.73 & 0.50 & 0.72 & 0.58 & 0.70 & 0.46 & 0.71 & 0.58 & 0.72 & 0.58 & 0.74 & 0.59 & 0.81 & 0.59 \\
\hline
 & 20 & 1 & 5 & 10 & 20 & 5.0 & 10.0 & 0.5 & 0.72 & 0.88 & 0.72 & 0.88 & 0.81 & 0.91 & 0.70 & 0.87 & 0.81 & 0.90 & 0.81 & 0.91 & 0.81 & 0.92 & 0.82 & 0.93 \\
 & 20 & 1 & 5 & 10 & 18 & 5.0 & 10.0 & 0.5 & 0.71 & 0.71 & 0.71 & 0.72 & 0.77 & 0.77 & 0.69 & 0.70 & 0.77 & 0.77 & 0.77 & 0.77 & 0.78 & 0.78 & 0.80 & 0.79 \\
 & 20 & 1 & 5 & 10 & 17 & 5.0 & 10.0 & 0.5 & 0.71 & 0.59 & 0.71 & 0.60 & 0.76 & 0.69 & 0.70 & 0.58 & 0.75 & 0.68 & 0.76 & 0.69 & 0.77 & 0.69 & 0.80 & 0.71 \\
 & 20 & 1 & 5 & 10 & 16 & 5.0 & 10.0 & 0.5 & 0.71 & 0.47 & 0.72 & 0.48 & 0.73 & 0.57 & 0.70 & 0.46 & 0.72 & 0.57 & 0.73 & 0.57 & 0.74 & 0.57 & 0.80 & 0.58 \\
\hline
 & 20 & 1 & 5 & 10 & 20 & 5.0 & 10.0 & 0.0 & 0.69 & 0.87 & 0.70 & 0.87 & 0.80 & 0.92 & 0.69 & 0.87 & 0.79 & 0.92 & 0.79 & 0.92 & 0.80 & 0.93 & 0.80 & 0.93 \\
 & 20 & 1 & 5 & 10 & 18 & 5.0 & 10.0 & 0.0 & 0.70 & 0.70 & 0.70 & 0.70 & 0.79 & 0.80 & 0.69 & 0.70 & 0.79 & 0.79 & 0.79 & 0.80 & 0.79 & 0.80 & 0.80 & 0.81 \\
 & 20 & 1 & 5 & 10 & 17 & 5.0 & 10.0 & 0.0 & 0.70 & 0.57 & 0.70 & 0.57 & 0.79 & 0.69 & 0.69 & 0.57 & 0.78 & 0.69 & 0.78 & 0.69 & 0.79 & 0.69 & 0.80 & 0.70 \\
\hline
 & 20 & 1 & 5 & 10 & 16 & 5.0 & 10.0 & 0.0 & 0.70 & 0.47 & 0.70 & 0.47 & 0.77 & 0.59 & 0.69 & 0.47 & 0.76 & 0.59 & 0.77 & 0.59 & 0.78 & 0.59 & 0.80 & 0.60 \\
 & 20 & 1 & 5 & 10 & 20 & 5.0 & 10.0 & -0.5 & 0.68 & 0.87 & 0.69 & 0.87 & 0.79 & 0.93 & 0.69 & 0.87 & 0.79 & 0.93 & 0.79 & 0.93 & 0.79 & 0.93 & 0.79 & 0.93 \\
 & 20 & 1 & 5 & 10 & 18 & 5.0 & 10.0 & -0.5 & 0.67 & 0.70 & 0.68 & 0.70 & 0.79 & 0.80 & 0.68 & 0.70 & 0.79 & 0.80 & 0.79 & 0.80 & 0.79 & 0.80 & 0.79 & 0.80 \\
 & 20 & 1 & 5 & 10 & 17 & 5.0 & 10.0 & -0.5 & 0.70 & 0.57 & 0.70 & 0.57 & 0.80 & 0.69 & 0.70 & 0.57 & 0.80 & 0.69 & 0.80 & 0.69 & 0.80 & 0.69 & 0.80 & 0.69 \\
 & 20 & 1 & 5 & 10 & 16 & 5.0 & 10.0 & -0.5 & 0.69 & 0.44 & 0.70 & 0.45 & 0.80 & 0.58 & 0.70 & 0.45 & 0.80 & 0.58 & 0.80 & 0.58 & 0.80 & 0.58 & 0.81 & 0.58 \\
\hline\hline
$H_{\delta,\delta}$ & 20 & 1 & 5 & 10 & 20 & 5.0 & 10.0 & 0.9 & 0.76 & 0.90 & 0.76 & 0.90 & 0.80 & 0.87 & 0.70 & 0.87 & 0.79 & 0.86 & 0.80 & 0.88 & 0.80 & 0.89 & 0.80 & 0.93 \\
 & 20 & 1 & 4 & 10 & 18 & 5.0 & 10.0 & 0.9 & 0.53 & 0.76 & 0.53 & 0.76 & 0.58 & 0.71 & 0.46 & 0.70 & 0.57 & 0.68 & 0.58 & 0.69 & 0.58 & 0.72 & 0.59 & 0.81 \\
 & 20 & 1 & 3 & 10 & 17 & 5.0 & 10.0 & 0.9 & 0.29 & 0.64 & 0.29 & 0.64 & 0.34 & 0.52 & 0.23 & 0.57 & 0.34 & 0.50 & 0.34 & 0.50 & 0.34 & 0.53 & 0.34 & 0.70 \\
 & 20 & 1 & 2 & 10 & 16 & 5.0 & 10.0 & 0.9 & 0.12 & 0.53 & 0.12 & 0.53 & 0.16 & 0.35 & 0.09 & 0.47 & 0.16 & 0.32 & 0.16 & 0.30 & 0.16 & 0.35 & 0.16 & 0.59 \\
 & 20 & 1 & 5 & 10 & 20 & 5.0 & 10.0 & 0.7 & 0.73 & 0.89 & 0.73 & 0.89 & 0.80 & 0.90 & 0.70 & 0.87 & 0.79 & 0.88 & 0.80 & 0.89 & 0.80 & 0.90 & 0.80 & 0.93 \\
 & 20 & 1 & 4 & 10 & 18 & 5.0 & 10.0 & 0.7 & 0.48 & 0.72 & 0.48 & 0.72 & 0.57 & 0.71 & 0.45 & 0.69 & 0.56 & 0.70 & 0.57 & 0.70 & 0.57 & 0.72 & 0.58 & 0.80 \\
 & 20 & 1 & 3 & 10 & 17 & 5.0 & 10.0 & 0.7 & 0.27 & 0.62 & 0.27 & 0.63 & 0.35 & 0.56 & 0.24 & 0.59 & 0.35 & 0.53 & 0.35 & 0.54 & 0.35 & 0.57 & 0.36 & 0.72 \\
 & 20 & 1 & 2 & 10 & 16 & 5.0 & 10.0 & 0.7 & 0.11 & 0.49 & 0.11 & 0.50 & 0.15 & 0.37 & 0.09 & 0.46 & 0.15 & 0.33 & 0.16 & 0.33 & 0.16 & 0.37 & 0.16 & 0.60 \\
 & 20 & 1 & 5 & 10 & 20 & 5.0 & 10.0 & 0.0 & 0.69 & 0.86 & 0.69 & 0.87 & 0.80 & 0.92 & 0.69 & 0.86 & 0.80 & 0.91 & 0.80 & 0.92 & 0.80 & 0.92 & 0.80 & 0.92 \\
 & 20 & 1 & 4 & 10 & 18 & 5.0 & 10.0 & 0.0 & 0.44 & 0.70 & 0.45 & 0.70 & 0.57 & 0.77 & 0.44 & 0.70 & 0.57 & 0.76 & 0.57 & 0.77 & 0.57 & 0.78 & 0.58 & 0.80 \\
 & 20 & 1 & 3 & 10 & 17 & 5.0 & 10.0 & 0.0 & 0.23 & 0.56 & 0.23 & 0.57 & 0.32 & 0.60 & 0.23 & 0.56 & 0.32 & 0.58 & 0.32 & 0.59 & 0.33 & 0.62 & 0.34 & 0.69 \\
 & 20 & 1 & 2 & 10 & 16 & 5.0 & 10.0 & 0.0 & 0.09 & 0.46 & 0.09 & 0.46 & 0.13 & 0.42 & 0.09 & 0.46 & 0.13 & 0.37 & 0.14 & 0.39 & 0.14 & 0.43 & 0.14 & 0.59 \\
 & 20 & 1 & 5 & 10 & 20 & 5.0 & 10.0 & -0.5 & 0.67 & 0.88 & 0.68 & 0.88 & 0.79 & 0.93 & 0.68 & 0.88 & 0.79 & 0.93 & 0.79 & 0.93 & 0.79 & 0.93 & 0.79 & 0.93 \\
 & 20 & 1 & 4 & 10 & 18 & 5.0 & 10.0 & -0.5 & 0.45 & 0.69 & 0.45 & 0.70 & 0.58 & 0.80 & 0.45 & 0.70 & 0.58 & 0.79 & 0.58 & 0.79 & 0.58 & 0.80 & 0.58 & 0.80 \\
 & 20 & 1 & 3 & 10 & 17 & 5.0 & 10.0 & -0.5 & 0.22 & 0.57 & 0.23 & 0.57 & 0.34 & 0.68 & 0.23 & 0.57 & 0.34 & 0.66 & 0.34 & 0.67 & 0.34 & 0.69 & 0.34 & 0.70 \\
 & 20 & 1 & 2 & 10 & 16 & 5.0 & 10.0 & -0.5 & 0.09 & 0.45 & 0.09 & 0.45 & 0.14 & 0.50 & 0.09 & 0.45 & 0.14 & 0.45 & 0.14 & 0.47 & 0.15 & 0.51 & 0.15 & 0.58 \\
   \hline
\end{tabular}
}
\end{table}

\subsubsection{Two sample design with 2 endpoints: all-pairs power}\label{ALL}
Table \ref{tab:all} shows all under $H_0$-estimates (under global $H_0$), the strong control of FWER (under partial $H_{01}$) 
and the all-pairs power for several noncentralities of both endpoints and $\rho=0.9,0.7,0.5,0, -0.5$.\\
Per definition, all multiple tests (maxT-tests, CTP, IUT) control the error rates in both weak and strong sense, with the remarkable tendency of conservatism towards low correlation. And that already with only 2 marginal tests, a basic property of the underlying IUT. 
 Per-definition, all tests become rather conservative under global $H_0$, particularly in direction of small $\rho$. 
The all-pairs power of the IUT is almost similar to those of the competitive maxT (or CTP), but the latter provide additional individual inference information (independent on $\rho$). Per definition is the power of the IUT and of the CTP identical under the all-pairs power concept.\\
For co-primary endpoints analysis, the maxT using the all-pairs power concept can be used as an alternative to the IUT when considering 2 correlated endpoints in a two-sample design, if one accepts a certain power loss in favor of the available individual inferences (i.e., simultaneous confidence intervals). 

\begin{table}[H]
\centering\footnotesize
\caption{Bivariate two-sample design: all-pairs power}
\label{tab:all}
\scalebox{0.5}{
\begin{tabular}{rrrrrrrrr|rr|rrrrr|r|rr}
  \hline
 & n & ma1 & ma2 & mb1 & mb2 & sa & sb & rho & maxT & mmm & CTPns & Bon & CTPd & CTPD & CTPps & IUT & Y1 & Y2 \\
  \hline
$H_0$ & 20 & 1 & 1 & 10 & 10 & 5.0 & 10.0 & 0.9 & 0.02 & 0.02 & 0.03 & 0.02 & 0.03 & 0.03 & 0.03 & 0.03 & 0.05 & 0.05 \\
 & 20 & 1 & 1 & 10 & 10 & 5.0 & 10.0 & 0.7 & 0.01 & 0.01 & 0.02 & 0.01 & 0.02 & 0.02 & 0.02 & 0.02 & 0.05 & 0.05 \\
 & 20 & 1 & 1 & 10 & 10 & 5.0 & 10.0 & 0.0 & 0.00 & 0.00 & 0.00 & 0.00 & 0.00 & 0.00 & 0.00 & 0.00 & 0.05 & 0.05 \\
 & 20 & 1 & 1 & 10 & 10 & 5.0 & 10.0 & -0.5 & 0.00 & 0.00 & 0.00 & 0.00 & 0.00 & 0.00 & 0.00 & 0.00 & 0.05 & 0.05 \\
\hline\hline	
$H_{0,1}$ & 20 & 1 & 1 & 10 & 20 & 5.0 & 10.0 & 0.9 & 0.02 & 0.02 & 0.08 & 0.02 & 0.08 & 0.08 & 0.08 & 0.08 & 0.08 & 0.84 \\
 & 20 & 1 & 1 & 10 & 18 & 5.0 & 10.0 & 0.9 & 0.03 & 0.03 & 0.04 & 0.02 & 0.04 & 0.04 & 0.04 & 0.04 & 0.04 & 0.80 \\
 & 20 & 1 & 1 & 10 & 17 & 5.0 & 10.0 & 0.9 & 0.04 & 0.04 & 0.05 & 0.03 & 0.05 & 0.05 & 0.05 & 0.05 & 0.05 & 0.71 \\
 & 20 & 1 & 1 & 10 & 16 & 5.0 & 10.0 & 0.9 & 0.03 & 0.03 & 0.05 & 0.02 & 0.05 & 0.05 & 0.05 & 0.05 & 0.05 & 0.58 \\
\hline 
& 20 & 1 & 1 & 10 & 20 & 5.0 & 10.0 & 0.7 & 0.03 & 0.03 & 0.05 & 0.02 & 0.05 & 0.05 & 0.05 & 0.05 & 0.05 & 0.93 \\
 & 20 & 1 & 1 & 10 & 18 & 5.0 & 10.0 & 0.7 & 0.03 & 0.03 & 0.05 & 0.03 & 0.05 & 0.05 & 0.05 & 0.05 & 0.05 & 0.80 \\
 & 20 & 1 & 1 & 10 & 17 & 5.0 & 10.0 & 0.7 & 0.03 & 0.03 & 0.05 & 0.02 & 0.05 & 0.05 & 0.05 & 0.05 & 0.05 & 0.70 \\
 & 20 & 1 & 1 & 10 & 16 & 5.0 & 10.0 & 0.7 & 0.03 & 0.03 & 0.05 & 0.02 & 0.05 & 0.05 & 0.05 & 0.05 & 0.05 & 0.60 \\
\hline 
& 20 & 1 & 1 & 10 & 20 & 5.0 & 10.0 & 0.5 & 0.03 & 0.03 & 0.05 & 0.03 & 0.05 & 0.05 & 0.05 & 0.05 & 0.05 & 0.93 \\
 & 20 & 1 & 1 & 10 & 18 & 5.0 & 10.0 & 0.5 & 0.03 & 0.03 & 0.05 & 0.02 & 0.05 & 0.05 & 0.05 & 0.05 & 0.05 & 0.80 \\
 & 20 & 1 & 1 & 10 & 17 & 5.0 & 10.0 & 0.5 & 0.02 & 0.02 & 0.04 & 0.02 & 0.04 & 0.04 & 0.04 & 0.04 & 0.05 & 0.71 \\
 & 20 & 1 & 1 & 10 & 16 & 5.0 & 10.0 & 0.5 & 0.02 & 0.02 & 0.05 & 0.02 & 0.05 & 0.05 & 0.05 & 0.05 & 0.05 & 0.58 \\
\hline 
& 20 & 1 & 1 & 10 & 20 & 5.0 & 10.0 & 0.0 & 0.02 & 0.02 & 0.05 & 0.02 & 0.05 & 0.05 & 0.05 & 0.05 & 0.05 & 0.93 \\
 & 20 & 1 & 1 & 10 & 18 & 5.0 & 10.0 & 0.0 & 0.02 & 0.02 & 0.04 & 0.02 & 0.04 & 0.04 & 0.04 & 0.04 & 0.05 & 0.80 \\
 & 20 & 1 & 1 & 10 & 17 & 5.0 & 10.0 & 0.0 & 0.02 & 0.02 & 0.04 & 0.02 & 0.04 & 0.04 & 0.04 & 0.04 & 0.05 & 0.71 \\
 & 20 & 1 & 1 & 10 & 16 & 5.0 & 10.0 & 0.0 & 0.01 & 0.01 & 0.03 & 0.01 & 0.03 & 0.03 & 0.03 & 0.03 & 0.05 & 0.59 \\
\hline \hline
$H_{1,1}$ & 20 & 1 & 5 & 10 & 20 & 5.0 & 10.0 & 0.9 & 0.75 & 0.75 & 0.79 & 0.69 & 0.79 & 0.80 & 0.80 & 0.80 & 0.80 & 0.93 \\
 & 20 & 1 & 5 & 10 & 18 & 5.0 & 10.0 & 0.9 & 0.70 & 0.70 & 0.74 & 0.63 & 0.73 & 0.74 & 0.75 & 0.75 & 0.80 & 0.80 \\
 & 20 & 1 & 5 & 10 & 17 & 5.0 & 10.0 & 0.9 & 0.62 & 0.62 & 0.67 & 0.54 & 0.66 & 0.68 & 0.68 & 0.68 & 0.80 & 0.70 \\
 & 20 & 1 & 5 & 10 & 16 & 5.0 & 10.0 & 0.9 & 0.52 & 0.52 & 0.58 & 0.45 & 0.56 & 0.58 & 0.58 & 0.58 & 0.80 & 0.59 \\
\hline 
& 20 & 1 & 5 & 10 & 20 & 5.0 & 10.0 & 0.7 & 0.70 & 0.70 & 0.78 & 0.66 & 0.78 & 0.78 & 0.78 & 0.78 & 0.80 & 0.93 \\
 & 20 & 1 & 5 & 10 & 18 & 5.0 & 10.0 & 0.7 & 0.62 & 0.63 & 0.72 & 0.58 & 0.71 & 0.72 & 0.72 & 0.72 & 0.80 & 0.81 \\
 & 20 & 1 & 5 & 10 & 17 & 5.0 & 10.0 & 0.7 & 0.53 & 0.54 & 0.63 & 0.50 & 0.63 & 0.63 & 0.63 & 0.63 & 0.80 & 0.69 \\
 & 20 & 1 & 5 & 10 & 16 & 5.0 & 10.0 & 0.7 & 0.45 & 0.45 & 0.55 & 0.42 & 0.54 & 0.55 & 0.55 & 0.55 & 0.80 & 0.58 \\
\hline 
& 20 & 1 & 5 & 10 & 20 & 5.0 & 10.0 & 0.5 & 0.67 & 0.67 & 0.77 & 0.65 & 0.77 & 0.77 & 0.77 & 0.77 & 0.81 & 0.93 \\
 & 20 & 1 & 5 & 10 & 18 & 5.0 & 10.0 & 0.5 & 0.57 & 0.57 & 0.70 & 0.54 & 0.69 & 0.70 & 0.70 & 0.70 & 0.81 & 0.80 \\
 & 20 & 1 & 5 & 10 & 17 & 5.0 & 10.0 & 0.5 & 0.50 & 0.51 & 0.62 & 0.49 & 0.62 & 0.62 & 0.62 & 0.62 & 0.80 & 0.70 \\
 & 20 & 1 & 5 & 10 & 16 & 5.0 & 10.0 & 0.5 & 0.40 & 0.41 & 0.52 & 0.39 & 0.52 & 0.52 & 0.52 & 0.52 & 0.80 & 0.58 \\
\hline 
& 20 & 1 & 5 & 10 & 20 & 5.0 & 10.0 & 0.0 & 0.60 & 0.60 & 0.74 & 0.60 & 0.74 & 0.74 & 0.74 & 0.74 & 0.80 & 0.93 \\
 & 20 & 1 & 5 & 10 & 18 & 5.0 & 10.0 & 0.0 & 0.48 & 0.48 & 0.63 & 0.48 & 0.63 & 0.63 & 0.63 & 0.63 & 0.80 & 0.79 \\
 & 20 & 1 & 5 & 10 & 17 & 5.0 & 10.0 & 0.0 & 0.41 & 0.41 & 0.56 & 0.41 & 0.56 & 0.56 & 0.56 & 0.56 & 0.80 & 0.71 \\
 & 20 & 1 & 5 & 10 & 16 & 5.0 & 10.0 & 0.0 & 0.31 & 0.31 & 0.46 & 0.30 & 0.46 & 0.46 & 0.46 & 0.46 & 0.80 & 0.58 \\
\hline 
& 20 & 1 & 5 & 10 & 20 & 5.0 & 10.0 & -0.5 & 0.57 & 0.57 & 0.73 & 0.57 & 0.73 & 0.73 & 0.73 & 0.73 & 0.80 & 0.93 \\
 & 20 & 1 & 5 & 10 & 18 & 5.0 & 10.0 & -0.5 & 0.42 & 0.42 & 0.60 & 0.42 & 0.60 & 0.60 & 0.60 & 0.60 & 0.80 & 0.80 \\
 & 20 & 1 & 5 & 10 & 17 & 5.0 & 10.0 & -0.5 & 0.33 & 0.34 & 0.52 & 0.34 & 0.52 & 0.52 & 0.52 & 0.52 & 0.79 & 0.71 \\
 & 20 & 1 & 5 & 10 & 16 & 5.0 & 10.0 & -0.5 & 0.24 & 0.25 & 0.41 & 0.25 & 0.41 & 0.41 & 0.41 & 0.41 & 0.81 & 0.57 \\
\hline\hline

$H_{\delta,\delta}$ & 20 & 1 & 5 & 10 & 20 & 5.0 & 10.0 & 0.9 & 0.75 & 0.75 & 0.80 & 0.69 & 0.79 & 0.80 & 0.80 & 0.80 & 0.81 & 0.93 \\
 & 20 & 1 & 4 & 10 & 18 & 5.0 & 10.0 & 0.9 & 0.52 & 0.52 & 0.57 & 0.45 & 0.57 & 0.58 & 0.58 & 0.58 & 0.58 & 0.80 \\
 & 20 & 1 & 3 & 10 & 17 & 5.0 & 10.0 & 0.9 & 0.29 & 0.29 & 0.34 & 0.23 & 0.34 & 0.34 & 0.34 & 0.34 & 0.35 & 0.71 \\
 & 20 & 1 & 2 & 10 & 16 & 5.0 & 10.0 & 0.9 & 0.12 & 0.12 & 0.15 & 0.09 & 0.15 & 0.15 & 0.15 & 0.15 & 0.15 & 0.58 \\
\hline 
& 20 & 1 & 5 & 10 & 20 & 5.0 & 10.0 & 0.7 & 0.70 & 0.70 & 0.78 & 0.67 & 0.78 & 0.78 & 0.78 & 0.78 & 0.80 & 0.92 \\
 & 20 & 1 & 4 & 10 & 18 & 5.0 & 10.0 & 0.7 & 0.44 & 0.44 & 0.54 & 0.41 & 0.54 & 0.54 & 0.55 & 0.55 & 0.59 & 0.80 \\
 & 20 & 1 & 3 & 10 & 17 & 5.0 & 10.0 & 0.7 & 0.23 & 0.24 & 0.32 & 0.21 & 0.32 & 0.32 & 0.32 & 0.32 & 0.34 & 0.70 \\
 & 20 & 1 & 2 & 10 & 16 & 5.0 & 10.0 & 0.7 & 0.11 & 0.11 & 0.15 & 0.09 & 0.15 & 0.15 & 0.15 & 0.15 & 0.16 & 0.60 \\
\hline 
& 20 & 1 & 5 & 10 & 20 & 5.0 & 10.0 & 0.0 & 0.61 & 0.61 & 0.75 & 0.61 & 0.75 & 0.75 & 0.75 & 0.75 & 0.80 & 0.93 \\
 & 20 & 1 & 4 & 10 & 18 & 5.0 & 10.0 & 0.0 & 0.31 & 0.31 & 0.47 & 0.31 & 0.47 & 0.47 & 0.47 & 0.47 & 0.59 & 0.80 \\
 & 20 & 1 & 3 & 10 & 17 & 5.0 & 10.0 & 0.0 & 0.13 & 0.13 & 0.23 & 0.13 & 0.23 & 0.23 & 0.23 & 0.23 & 0.33 & 0.70 \\
 & 20 & 1 & 2 & 10 & 16 & 5.0 & 10.0 & 0.0 & 0.04 & 0.04 & 0.09 & 0.04 & 0.09 & 0.09 & 0.09 & 0.09 & 0.15 & 0.57 \\
\hline
 & 20 & 1 & 5 & 10 & 20 & 5.0 & 10.0 & -0.5 & 0.56 & 0.57 & 0.73 & 0.57 & 0.73 & 0.73 & 0.73 & 0.73 & 0.80 & 0.93 \\
 & 20 & 1 & 4 & 10 & 18 & 5.0 & 10.0 & -0.5 & 0.24 & 0.24 & 0.41 & 0.24 & 0.41 & 0.41 & 0.41 & 0.41 & 0.58 & 0.81 \\
 & 20 & 1 & 3 & 10 & 17 & 5.0 & 10.0 & -0.5 & 0.07 & 0.08 & 0.18 & 0.08 & 0.18 & 0.18 & 0.18 & 0.18 & 0.35 & 0.70 \\
 & 20 & 1 & 2 & 10 & 16 & 5.0 & 10.0 & -0.5 & 0.01 & 0.01 & 0.04 & 0.01 & 0.04 & 0.04 & 0.04 & 0.04 & 0.15 & 0.58 \\
   \hline
\end{tabular}
}
\end{table}

\subsection{Two sample design with 3 correlated endpoints}
\subsubsection{Two sample design with 3 endpoints: any-pairs power}
Compared to the simulation results with 2 endpoints, for 3 endpoints similar $H_0$ properties occur in Table \ref{tab:threeany}. Again, under specific conditions the any-pairs power may be larger than the comparisonwise power. The summarzing global tests within the CTP (OBrien, distance based tests) are only powerful when all endpoints contribute to $H_1$ with some noncentrality- in direction of lower correlations- and vice versa. The summarzing tests reveal somewhat higher  power than max-T tests only in direction of lower correlations. As long the correlations are not towards to zero, the Bonferroni test is acceptable. 

\begin{table}[H]
\centering
\caption{Tri-variate two-sample design: any-pairs power}
\label{tab:threeany}
\scalebox{0.5}{
\begin{tabular}{rrrrrrrrrrr|rrrrrr|rrr}
  \hline
 & n & ma1 & ma2 & mb1 & mb2 & sa & sb & rho12 & rho13 & rho23 & maxT & mmm & Bon & CTPd & CTPD & CTPps & Y1 & Y2 & Y3 \\ 
  \hline
$H_0$	 	& 20 & 1 & 1 & 2 & 2 & 1.0 & 4.0 & 0.3 & 0.5 & 0.7 & 0.05 & 0.05 & 0.04 & 0.03 & 0.03 & 0.04 & 0.05 & 0.05 & 0.05 \\ 
				& 20 & 1 & 1 & 2 & 2 & 1.0 & 4.0 & 0.0 & 0.1 & 0.5 & 0.06 & 0.06 & 0.05 & 0.02 & 0.02 & 0.03 & 0.05 & 0.05 & 0.06 \\ 
				& 20 & 1 & 1 & 2 & 2 & 1.0 & 4.0 & 0.2 & 0.5 & 0.9 & 0.05 & 0.05 & 0.04 & 0.03 & 0.03 & 0.03 & 0.05 & 0.05 & 0.05 \\ 
				& 20 & 1 & 1 & 2 & 2 & 1.0 & 4.0 & 0.0 & 0.0 & 0.0 & 0.05 & 0.05 & 0.05 & 0.01 & 0.02 & 0.02 & 0.05 & 0.05 & 0.05 \\ 
\hline				
$H_{011}$ & 20 & 1 & 1 & 2 & 5 & 1.0 & 4.0 & 0.2 & 0.5 & 0.9 & 0.76 & 0.76 & 0.72 & 0.34 & 0.35 & 0.46 & 0.06 & 0.76 & 0.72 \\ 
					& 20 & 1 & 1 & 2 & 5 & 1.0 & 4.0 & 0.0 & 0.0 & 0.0 & 0.78 & 0.79 & 0.78 & 0.41 & 0.44 & 0.56 & 0.05 & 0.74 & 0.72 \\ 
					& 20 & 1 & 1 & 2 & 5 & 1.0 & 4.0 & 0.0 & 0.1 & 0.5 & 0.79 & 0.80 & 0.78 & 0.38 & 0.43 & 0.51 & 0.05 & 0.76 & 0.74 \\ 
					& 20 & 1 & 1 & 2 & 5 & 1.0 & 4.0 & 0.3 & 0.5 & 0.7 & 0.77 & 0.77 & 0.73 & 0.34 & 0.34 & 0.44 & 0.05 & 0.75 & 0.74 \\ 
   \hline
$H_{111}$ & 20 & 1 & 2 & 2 & 5 & 1.0 & 4.0 & 0.3 & 0.5 & 0.7 & 0.81 & 0.81 & 0.78 & 0.78 & 0.78 & 0.82 & 0.57 & 0.76 & 0.74 \\ 
 & 20 & 1 & 2 & 2 & 5 & 1.0 & 4.0 & 0.0 & 0.0 & 0.0 & 0.88 & 0.88 & 0.88 & 0.89 & 0.91 & 0.94 & 0.59 & 0.76 & 0.72 \\ 
 & 20 & 1 & 2 & 2 & 5 & 1.0 & 4.0 & 0.0 & 0.1 & 0.5 & 0.86 & 0.86 & 0.85 & 0.85 & 0.87 & 0.89 & 0.57 & 0.77 & 0.74 \\ 
 & 20 & 1 & 2 & 2 & 5 & 1.0 & 4.0 & 0.2 & 0.5 & 0.9 & 0.79 & 0.79 & 0.74 & 0.76 & 0.80 & 0.81 & 0.58 & 0.76 & 0.71 \\ 
 & 20 & 1 & 2 & 2 & 5 & 1.0 & 4.0 & 0.9 & 0.9 & 0.9 & 0.72 & 0.72 & 0.64 & 0.66 & 0.66 & 0.69 & 0.58 & 0.76 & 0.72 \\ 
    \hline
\end{tabular}
}
\end{table}

\subsubsection{Two sample design with 3 endpoints: all-pairs power}
Table \ref{tab:3all} shows the strong conservative properties of all tests. The all-pairs power of the IUT is higher than those of the competitive maxT or CTP.

\begin{table}[H]
\caption{Tri-variate two-sample design: all-pairs power}
\label{tab:3all}
\centering
\scalebox{0.5}{
\begin{tabular}{rrrrrrrrrrrrrrrrrrrrr}
  \hline
 & n & ma1 & ma2 & mb1 & mb2 & sa & sb & rho1 & rho2 & rho3 & maxT & mmm & Bon & CTPd & CTPD & CTPps & IUT & Y1 & Y2 & Y3 \\ 
  \hline
$H_{111}$ & 20 & 1 & 2 & 2 & 5 & 1.0 & 4.0 & 0.9 & 0.9 & 0.9 & 0.46 & 0.46 & 0.34 & 0.54 & 0.56 & 0.56 & 0.56 & 0.58 & 0.74 & 0.73 \\ 
 & 20 & 1 & 2 & 2 & 5 & 1.0 & 4.0 & 0.3 & 0.5 & 0.7 & 0.27 & 0.28 & 0.24 & 0.42 & 0.42 & 0.42 & 0.42 & 0.56 & 0.72 & 0.74 \\ 
 & 20 & 1 & 2 & 2 & 5 & 1.0 & 4.0 & 0.0 & 0.0 & 0.0 & 0.13 & 0.13 & 0.13 & 0.31 & 0.31 & 0.31 & 0.31 & 0.57 & 0.74 & 0.73 \\ 
 & 20 & 1 & 2 & 2 & 5 & 1.0 & 4.0 & 0.0 & 0.1 & 0.5 & 0.17 & 0.17 & 0.16 & 0.40 & 0.40 & 0.40 & 0.40 & 0.60 & 0.75 & 0.74 \\ 
 & 20 & 1 & 2 & 2 & 5 & 1.0 & 4.0 & 0.2 & 0.5 & 0.9 & 0.30 & 0.30 & 0.24 & 0.45 & 0.46 & 0.46 & 0.46 & 0.60 & 0.75 & 0.73 \\ 
\hline
$H_{000}$ & 20 & 1 & 1 & 2 & 2 & 1.0 & 4.0 & 0.3 & 0.5 & 0.7 & 0.00 & 0.00 & 0.00 & 0.01 & 0.01 & 0.01 & 0.01 & 0.04 & 0.07 & 0.06 \\ 
 & 20 & 1 & 1 & 2 & 2 & 1.0 & 4.0 & 0.0 & 0.0 & 0.0 & 0.00 & 0.00 & 0.00 & 0.00 & 0.00 & 0.00 & 0.00 & 0.06 & 0.05 & 0.05 \\ 
 & 20 & 1 & 1 & 2 & 2 & 1.0 & 4.0 & 0.0 & 0.1 & 0.5 & 0.00 & 0.00 & 0.00 & 0.00 & 0.00 & 0.00 & 0.00 & 0.06 & 0.05 & 0.05 \\ 
 & 20 & 1 & 1 & 2 & 2 & 1.0 & 4.0 & 0.2 & 0.5 & 0.9 & 0.00 & 0.00 & 0.00 & 0.00 & 0.01 & 0.01 & 0.01 & 0.05 & 0.05 & 0.06 \\ 
   \hline
$H_{011}$ & 20 & 1 & 1 & 2 & 5 & 1.0 & 4.0 & 0.9 & 0.9 & 0.9 & 0.03 & 0.03 & 0.01 & 0.04 & 0.04 & 0.04 & 0.04 & 0.04 & 0.74 & 0.71 \\ 
 & 20 & 1 & 1 & 2 & 5 & 1.0 & 4.0 & 0.3 & 0.5 & 0.7 & 0.02 & 0.02 & 0.02 & 0.05 & 0.05 & 0.05 & 0.05 & 0.06 & 0.74 & 0.73 \\ 
 & 20 & 1 & 1 & 2 & 5 & 1.0 & 4.0 & 0.0 & 0.0 & 0.0 & 0.00 & 0.00 & 0.00 & 0.03 & 0.03 & 0.03 & 0.03 & 0.05 & 0.75 & 0.74 \\ 
 & 20 & 1 & 1 & 2 & 5 & 1.0 & 4.0 & 0.0 & 0.1 & 0.5 & 0.01 & 0.01 & 0.01 & 0.03 & 0.03 & 0.03 & 0.03 & 0.04 & 0.75 & 0.73 \\ 
 & 20 & 1 & 1 & 2 & 5 & 1.0 & 4.0 & 0.2 & 0.5 & 0.9 & 0.01 & 0.01 & 0.01 & 0.03 & 0.03 & 0.03 & 0.03 & 0.04 & 0.77 & 0.74 \\ 
   \hline
\end{tabular}
}
\end{table}

\subsection{Bivariate Dunnett-Type comparisons}
As a simple example for multivariate Dunnett-type comparisons, the bivariate design with 2 endpoints is considered. The question arises which impact have two sources of multiplicity, namely between treatment-versus-control comparisons and between correlate endpoints.

\subsubsection{Bivariate Dunnett-type test: global and any pairs power}
All procedures controls the familywise error rate in both weak and strong sense as expected, see Table \ref{tab:duany}.
The power estimated of maxT and mmm-test are always the same under the considered configurations. 
The power of the global CTPd test (ie Euclidean distance global test) (gd) is smaller than CTPD (i.e. maximum absolute component distance global test) (gD). The power of both distance-based global tests (gD,gd) is smaller than of the parametric OBrien-type summary test(gps).
Only under selected circumstances, particularly of near-to-zero correlation,  the any-pairs power can be slightly higher than any of the comparisonwise power (based on CWER control only).  The power behavior of all tests in the bivariate Dunnett-type design is similar behavior to those in the two-sample design. The power of all summarzing tests is lower when any treatment does not contribute to $H_1$.

\begin{table}[H]
\caption{Bivariate Dunnett-type comparisons: any-pairs and global power}
\centering
\label{tab:duany}
\scalebox{0.5}{
\begin{tabular}{rrrrrrrrrrr|rr|rrrr||rrr||rrrr}
 & n & ma1 & ma2 & ma3 & mb1 & mb2 & mb3 & sa & sb & rho & maxT & mmm & CTPns & Bon & CTPd & CPTD & gd & gD & gps & Y11 & Y12 & Y21 & Y22 \\
  \hline
$H_0$ & 20 & 1 & 1 & 1 & 10 & 10 & 10 & 5.0 & 10.0 & 0.9 & 0.06 & 0.05 & 0.05 & 0.04 & 0.03 & 0.05 & 0.05 & 0.05 & 0.05 & 0.05 & 0.05 & 0.05 & 0.05 \\
 & 20 & 1 & 1 & 1 & 10 & 10 & 10 & 5.0 & 10.0 & 0.7 & 0.05 & 0.05 & 0.04 & 0.04 & 0.03 & 0.05 & 0.05 & 0.05 & 0.05 & 0.05 & 0.05 & 0.05 & 0.05 \\
 & 20 & 1 & 1 & 1 & 10 & 10 & 10 & 5.0 & 10.0 & 0.5 & 0.05 & 0.05 & 0.03 & 0.04 & 0.02 & 0.04 & 0.04 & 0.04 & 0.05 & 0.05 & 0.05 & 0.04 & 0.05 \\
 & 20 & 1 & 1 & 1 & 10 & 10 & 10 & 5.0 & 10.0 & 0.1 & 0.06 & 0.06 & 0.02 & 0.05 & 0.02 & 0.05 & 0.05 & 0.05 & 0.06 & 0.05 & 0.05 & 0.06 & 0.05 \\
 & 20 & 1 & 1 & 1 & 10 & 10 & 10 & 5.0 & 10.0 & -0.5 & 0.05 & 0.06 & 0.01 & 0.05 & 0.00 & 0.04 & 0.04 & 0.04 & 0.05 & 0.05 & 0.05 & 0.05 & 0.05 \\
   \hline
$H_{1111}$ & 20 & 1 & 2 & 5 & 10 & 15 & 20 & 5.0 & 10.0 & 0.9 & 0.85 & 0.85 & 0.81 & 0.80 & 0.74 & 0.77 & 0.76 & 0.80 & 0.83 & 0.15 & 0.80 & 0.45 & 0.92 \\
   \hline
 & 20 & 1 & 2 & 5 & 10 & 15 & 20 & 5.0 & 10.0 & 0.7 & 0.86 & 0.86 & 0.81 & 0.83 & 0.75 & 0.77 & 0.79 & 0.81 & 0.86 & 0.12 & 0.78 & 0.45 & 0.92 \\
 & 20 & 1 & 2 & 5 & 10 & 15 & 20 & 5.0 & 10.0 & 0.5 & 0.88 & 0.89 & 0.83 & 0.87 & 0.79 & 0.81 & 0.85 & 0.87 & 0.90 & 0.15 & 0.80 & 0.46 & 0.92 \\
 & 20 & 1 & 2 & 5 & 10 & 15 & 20 & 5.0 & 10.0 & 0.0 & 0.93 & 0.94 & 0.89 & 0.93 & 0.86 & 0.87 & 0.94 & 0.97 & 0.98 & 0.15 & 0.81 & 0.47 & 0.94 \\
\hline
 & 20 & 1 & 4 & 5 & 10 & 15 & 20 & 5.0 & 10.0 & 0.9 & 0.87 & 0.87 & 0.84 & 0.83 & 0.77 & 0.80 & 0.79 & 0.81 & 0.84 & 0.46 & 0.79 & 0.47 & 0.94 \\
 & 20 & 1 & 4 & 5 & 10 & 15 & 20 & 5.0 & 10.0 & 0.7 & 0.87 & 0.88 & 0.85 & 0.85 & 0.79 & 0.82 & 0.81 & 0.84 & 0.87 & 0.45 & 0.79 & 0.47 & 0.93 \\
 & 20 & 1 & 4 & 5 & 10 & 15 & 20 & 5.0 & 10.0 & 0.5 & 0.90 & 0.90 & 0.88 & 0.89 & 0.84 & 0.85 & 0.88 & 0.89 & 0.92 & 0.48 & 0.80 & 0.45 & 0.94 \\
 & 20 & 1 & 4 & 5 & 10 & 15 & 20 & 5.0 & 10.0 & 0.0 & 0.93 & 0.94 & 0.92 & 0.93 & 0.90 & 0.91 & 0.95 & 0.97 & 0.98 & 0.50 & 0.81 & 0.46 & 0.93 \\
\hline
 & 20 & 1 & 1 & 5 & 10 & 15 & 20 & 5.0 & 10.0 & 0.9 & 0.87 & 0.87 & 0.80 & 0.82 & 0.73 & 0.76 & 0.78 & 0.81 & 0.84 & 0.05 & 0.80 & 0.45 & 0.94 \\
 & 20 & 1 & 1 & 5 & 10 & 15 & 20 & 5.0 & 10.0 & 0.7 & 0.85 & 0.86 & 0.78 & 0.82 & 0.74 & 0.76 & 0.80 & 0.82 & 0.85 & 0.06 & 0.79 & 0.47 & 0.92 \\
 & 20 & 1 & 1 & 5 & 10 & 15 & 20 & 5.0 & 10.0 & 0.5 & 0.87 & 0.87 & 0.80 & 0.85 & 0.76 & 0.79 & 0.84 & 0.87 & 0.89 & 0.04 & 0.80 & 0.46 & 0.94 \\
 & 20 & 1 & 1 & 5 & 10 & 15 & 20 & 5.0 & 10.0 & 0.0 & 0.94 & 0.94 & 0.83 & 0.94 & 0.81 & 0.82 & 0.94 & 0.96 & 0.97 & 0.05 & 0.81 & 0.47 & 0.93 \\
\hline
 & 20 & 1 & 1 & 1 & 10 & 15 & 20 & 5.0 & 10.0 & 0.9 & 0.87 & 0.87 & 0.37 & 0.83 & 0.24 & 0.22 & 0.26 & 0.23 & 0.37 & 0.05 & 0.05 & 0.48 & 0.94 \\
 & 20 & 1 & 1 & 1 & 10 & 15 & 20 & 5.0 & 10.0 & 0.7 & 0.84 & 0.84 & 0.41 & 0.80 & 0.26 & 0.26 & 0.27 & 0.27 & 0.41 & 0.05 & 0.04 & 0.48 & 0.92 \\
 & 20 & 1 & 1 & 1 & 10 & 15 & 20 & 5.0 & 10.0 & 0.5 & 0.81 & 0.81 & 0.43 & 0.80 & 0.27 & 0.28 & 0.29 & 0.30 & 0.44 & 0.05 & 0.04 & 0.47 & 0.92 \\
 & 20 & 1 & 1 & 1 & 10 & 15 & 20 & 5.0 & 10.0 & 0.0 & 0.81 & 0.82 & 0.57 & 0.81 & 0.37 & 0.39 & 0.42 & 0.45 & 0.61 & 0.05 & 0.06 & 0.47 & 0.94 \\
\hline

 & 20 & 1 & 4 & 5 & 10 & 13 & 20 & 5.0 & 10.0 & 0.9 & 0.87 & 0.88 & 0.84 & 0.83 & 0.76 & 0.80 & 0.78 & 0.81 & 0.84 & 0.49 & 0.80 & 0.26 & 0.95 \\
 & 20 & 1 & 4 & 5 & 10 & 13 & 20 & 5.0 & 10.0 & 0.7 & 0.88 & 0.88 & 0.86 & 0.85 & 0.81 & 0.84 & 0.83 & 0.85 & 0.88 & 0.46 & 0.82 & 0.25 & 0.93 \\
 & 20 & 1 & 4 & 5 & 10 & 13 & 20 & 5.0 & 10.0 & 0.5 & 0.89 & 0.90 & 0.88 & 0.88 & 0.83 & 0.85 & 0.87 & 0.88 & 0.92 & 0.48 & 0.80 & 0.27 & 0.93 \\
 & 20 & 1 & 4 & 5 & 10 & 13 & 20 & 5.0 & 10.0 & 0.0 & 0.93 & 0.93 & 0.92 & 0.92 & 0.89 & 0.90 & 0.95 & 0.96 & 0.98 & 0.48 & 0.80 & 0.26 & 0.92 \\
 \hline\hline
$H_{1111}$ & 5 & 1 & 2 & 5 & 10 & 15 & 20 & 5.0 & 10.0 & 0.9 & 0.29 & 0.32 & 0.27 & 0.25 & 0.20 & 0.22 & 0.27 & 0.28 & 0.30 & 0.08 & 0.32 & 0.17 & 0.42 \\ 
 & 5 & 1 & 2 & 5 & 10 & 15 & 20 & 5.0 & 10.0 & 0.7 & 0.29 & 0.34 & 0.28 & 0.28 & 0.23 & 0.24 & 0.30 & 0.30 & 0.34 & 0.10 & 0.32 & 0.19 & 0.43 \\ 
 & 5 & 1 & 2 & 5 & 10 & 15 & 20 & 5.0 & 10.0 & 0.5 & 0.28 & 0.35 & 0.27 & 0.31 & 0.21 & 0.22 & 0.32 & 0.31 & 0.35 & 0.08 & 0.32 & 0.19 & 0.43 \\ 
 & 5 & 1 & 2 & 5 & 10 & 15 & 20 & 5.0 & 10.0 & 0.0 & 0.27 & 0.36 & 0.26 & 0.32 & 0.20 & 0.21 & 0.37 & 0.39 & 0.47 & 0.09 & 0.32 & 0.17 & 0.44 \\ \hline
 & 5 & 1 & 4 & 5 & 10 & 15 & 20 & 5.0 & 10.0 & 0.9 & 0.29 & 0.36 & 0.31 & 0.27 & 0.23 & 0.24 & 0.27 & 0.28 & 0.33 & 0.18 & 0.31 & 0.19 & 0.44 \\ 
 & 5 & 1 & 4 & 5 & 10 & 15 & 20 & 5.0 & 10.0 & 0.7 & 0.28 & 0.34 & 0.31 & 0.28 & 0.25 & 0.26 & 0.32 & 0.33 & 0.35 & 0.20 & 0.32 & 0.19 & 0.45 \\ 

 & 5 & 1 & 4 & 5 & 10 & 15 & 20 & 5.0 & 10.0 & 0.5 & 0.32 & 0.38 & 0.33 & 0.33 & 0.28 & 0.28 & 0.35 & 0.36 & 0.41 & 0.21 & 0.33 & 0.20 & 0.46 \\ 
 & 5 & 1 & 4 & 5 & 10 & 15 & 20 & 5.0 & 10.0 & 0.0 & 0.28 & 0.39 & 0.31 & 0.35 & 0.24 & 0.25 & 0.42 & 0.43 & 0.49 & 0.16 & 0.33 & 0.20 & 0.45 \\ \hline

 & 5 & 1 & 1 & 5 & 10 & 15 & 20 & 5.0 & 10.0 & 0.9 & 0.28 & 0.32 & 0.26 & 0.24 & 0.20 & 0.21 & 0.25 & 0.27 & 0.30 & 0.04 & 0.31 & 0.19 & 0.43 \\ 
 & 5 & 1 & 1 & 5 & 10 & 15 & 20 & 5.0 & 10.0 & 0.7 & 0.29 & 0.34 & 0.26 & 0.28 & 0.20 & 0.21 & 0.28 & 0.31 & 0.34 & 0.04 & 0.34 & 0.18 & 0.42 \\ 
 & 5 & 1 & 1 & 5 & 10 & 15 & 20 & 5.0 & 10.0 & 0.5 & 0.31 & 0.35 & 0.25 & 0.31 & 0.20 & 0.21 & 0.32 & 0.34 & 0.38 & 0.05 & 0.33 & 0.20 & 0.46 \\ 
 & 5 & 1 & 1 & 5 & 10 & 15 & 20 & 5.0 & 10.0 & 0.0 & 0.28 & 0.37 & 0.24 & 0.34 & 0.18 & 0.19 & 0.39 & 0.42 & 0.47 & 0.06 & 0.33 & 0.20 & 0.45 \\ \hline
 & 5 & 1 & 1 & 1 & 10 & 15 & 20 & 5.0 & 10.0 & 0.9 & 0.26 & 0.31 & 0.15 & 0.24 & 0.10 & 0.10 & 0.12 & 0.12 & 0.15 & 0.06 & 0.06 & 0.20 & 0.46 \\ 
 & 5 & 1 & 1 & 1 & 10 & 15 & 20 & 5.0 & 10.0 & 0.7 & 0.26 & 0.30 & 0.14 & 0.25 & 0.09 & 0.10 & 0.11 & 0.11 & 0.15 & 0.05 & 0.05 & 0.19 & 0.45 \\ 
 & 5 & 1 & 1 & 1 & 10 & 15 & 20 & 5.0 & 10.0 & 0.5 & 0.21 & 0.27 & 0.15 & 0.24 & 0.10 & 0.10 & 0.13 & 0.15 & 0.17 & 0.05 & 0.05 & 0.18 & 0.43 \\ 
 & 5 & 1 & 1 & 1 & 10 & 15 & 20 & 5.0 & 10.0 & 0.0 & 0.22 & 0.28 & 0.14 & 0.25 & 0.07 & 0.08 & 0.16 & 0.16 & 0.21 & 0.05 & 0.04 & 0.22 & 0.44 \\  \hline
 & 5 & 1 & 4 & 5 & 10 & 13 & 20 & 5.0 & 10.0 & 0.9 & 0.29 & 0.33 & 0.29 & 0.26 & 0.22 & 0.24 & 0.27 & 0.28 & 0.30 & 0.17 & 0.30 & 0.12 & 0.43 \\ 
 & 5 & 1 & 4 & 5 & 10 & 13 & 20 & 5.0 & 10.0 & 0.7 & 0.29 & 0.37 & 0.31 & 0.30 & 0.22 & 0.23 & 0.29 & 0.30 & 0.36 & 0.20 & 0.34 & 0.12 & 0.46 \\ 
 & 5 & 1 & 4 & 5 & 10 & 13 & 20 & 5.0 & 10.0 & 0.5 & 0.29 & 0.36 & 0.29 & 0.30 & 0.23 & 0.24 & 0.32 & 0.33 & 0.37 & 0.18 & 0.32 & 0.13 & 0.46 \\ 
 & 5 & 1 & 4 & 5 & 10 & 13 & 20 & 5.0 & 10.0 & 0.0 & 0.32 & 0.39 & 0.29 & 0.36 & 0.22 & 0.23 & 0.42 & 0.42 & 0.49 & 0.20 & 0.33 & 0.12 & 0.44 \\    \hline

\end{tabular}
}
\end{table}

\subsubsection{Bivariate Dunnett-type test: individual power}

Table \ref{tab:duind} reveals as expected, each individual power in each test considered is lower than the corresponding comparisonwise power, i.e. in each case a multiplicity price is paid. 

\begin{table}[H]
\caption{Bivariate Dunnett-type comparisons: individual power}
\centering
\label{tab:duind}
\scalebox{0.273}{
\begin{tabular}{rrrrrrrrrrr|rrrr|rrrr|rrrr|rrrr|rrrr|rrrr|rrrr}
  \hline
 & n & ma1 & ma2 & ma3 & mb1 & mb2 & mb3 & sa & sb & rho & maxT11 & maxT12 & maxT21 & maxT22 & mmm11 & mmm12 & mmm21 & mmm22 & CTPps11 & CTPps12 & CTPps21 & CTPps22 & Bon11 & Bon12 & Bon21 & Bon22 & CTPd11 & CTPd12 & CTPd21 & CTPd22 & CTPD11 & CTPD12 & CTPD21 & CTPD22 & Y11 & Y12 & Y21 & Y22 \\ 
  \hline
$H_{1111}$ & 20 & 1 & 2 & 5 & 10 & 15 & 20 & 5.0 & 10.0 & 0.9 & 0.07 & 0.29 & 0.63 & 0.84 & 0.07 & 0.29 & 0.64 & 0.85 & 0.14 & 0.28 & 0.37 & 0.82 & 0.05 & 0.23 & 0.57 & 0.80 & 0.13 & 0.26 & 0.36 & 0.73 & 0.13 & 0.24 & 0.36 & 0.76 & 0.14 & 0.45 & 0.80 & 0.93 \\ 
 & 20 & 1 & 2 & 5 & 10 & 15 & 20 & 5.0 & 10.0 & 0.7 & 0.06 & 0.29 & 0.60 & 0.82 & 0.06 & 0.28 & 0.62 & 0.82 & 0.14 & 0.32 & 0.39 & 0.83 & 0.05 & 0.25 & 0.57 & 0.79 & 0.13 & 0.28 & 0.39 & 0.75 & 0.14 & 0.29 & 0.39 & 0.78 & 0.16 & 0.48 & 0.81 & 0.93 \\ 
 & 20 & 1 & 2 & 5 & 10 & 15 & 20 & 5.0 & 10.0 & 0.5 & 0.06 & 0.25 & 0.60 & 0.81 & 0.06 & 0.26 & 0.61 & 0.82 & 0.14 & 0.29 & 0.38 & 0.84 & 0.05 & 0.23 & 0.58 & 0.80 & 0.13 & 0.27 & 0.37 & 0.81 & 0.13 & 0.27 & 0.37 & 0.83 & 0.15 & 0.47 & 0.81 & 0.93 \\ 
 & 20 & 1 & 2 & 5 & 10 & 15 & 20 & 5.0 & 10.0 & 0.0 & 0.06 & 0.26 & 0.61 & 0.83 & 0.06 & 0.29 & 0.62 & 0.84 & 0.12 & 0.33 & 0.38 & 0.87 & 0.06 & 0.27 & 0.61 & 0.82 & 0.12 & 0.29 & 0.38 & 0.85 & 0.12 & 0.30 & 0.38 & 0.86 & 0.14 & 0.49 & 0.81 & 0.94 \\ 
\hline
 & 20 & 1 & 4 & 5 & 10 & 15 & 20 & 5.0 & 10.0 & 0.9 & 0.30 & 0.30 & 0.65 & 0.86 & 0.30 & 0.31 & 0.68 & 0.85 & 0.42 & 0.43 & 0.50 & 0.84 & 0.25 & 0.25 & 0.59 & 0.81 & 0.39 & 0.39 & 0.48 & 0.76 & 0.40 & 0.40 & 0.49 & 0.78 & 0.47 & 0.47 & 0.80 & 0.93 \\ 
 & 20 & 1 & 4 & 5 & 10 & 15 & 20 & 5.0 & 10.0 & 0.7 & 0.29 & 0.30 & 0.64 & 0.84 & 0.30 & 0.31 & 0.64 & 0.85 & 0.43 & 0.44 & 0.55 & 0.86 & 0.26 & 0.28 & 0.60 & 0.81 & 0.40 & 0.42 & 0.54 & 0.80 & 0.41 & 0.43 & 0.54 & 0.83 & 0.49 & 0.49 & 0.80 & 0.95 \\ 
 & 20 & 1 & 4 & 5 & 10 & 15 & 20 & 5.0 & 10.0 & 0.5 & 0.24 & 0.24 & 0.60 & 0.81 & 0.25 & 0.25 & 0.62 & 0.81 & 0.39 & 0.40 & 0.51 & 0.84 & 0.23 & 0.23 & 0.59 & 0.79 & 0.36 & 0.38 & 0.51 & 0.79 & 0.37 & 0.39 & 0.51 & 0.82 & 0.46 & 0.46 & 0.78 & 0.93 \\ 
 & 20 & 1 & 4 & 5 & 10 & 15 & 20 & 5.0 & 10.0 & 0.0 & 0.21 & 0.26 & 0.59 & 0.80 & 0.22 & 0.27 & 0.59 & 0.82 & 0.38 & 0.40 & 0.54 & 0.88 & 0.21 & 0.26 & 0.57 & 0.80 & 0.36 & 0.38 & 0.54 & 0.87 & 0.37 & 0.39 & 0.54 & 0.87 & 0.44 & 0.46 & 0.80 & 0.94 \\ 
\hline
 & 20 & 1 & 1 & 5 & 10 & 15 & 20 & 5.0 & 10.0 & 0.9 & 0.02 & 0.28 & 0.68 & 0.87 & 0.02 & 0.29 & 0.69 & 0.88 & 0.06 & 0.17 & 0.29 & 0.82 & 0.02 & 0.23 & 0.62 & 0.83 & 0.06 & 0.15 & 0.29 & 0.76 & 0.06 & 0.14 & 0.29 & 0.79 & 0.06 & 0.46 & 0.82 & 0.93 \\ 
 & 20 & 1 & 1 & 5 & 10 & 15 & 20 & 5.0 & 10.0 & 0.7 & 0.02 & 0.29 & 0.62 & 0.82 & 0.01 & 0.29 & 0.63 & 0.83 & 0.05 & 0.21 & 0.30 & 0.80 & 0.01 & 0.25 & 0.59 & 0.80 & 0.05 & 0.18 & 0.30 & 0.74 & 0.05 & 0.17 & 0.30 & 0.78 & 0.05 & 0.49 & 0.80 & 0.93 \\ 
 & 20 & 1 & 1 & 5 & 10 & 15 & 20 & 5.0 & 10.0 & 0.5 & 0.01 & 0.26 & 0.61 & 0.79 & 0.02 & 0.26 & 0.61 & 0.81 & 0.05 & 0.21 & 0.29 & 0.80 & 0.02 & 0.24 & 0.58 & 0.79 & 0.05 & 0.18 & 0.29 & 0.77 & 0.04 & 0.16 & 0.29 & 0.78 & 0.05 & 0.45 & 0.81 & 0.92 \\ 
 & 20 & 1 & 1 & 5 & 10 & 15 & 20 & 5.0 & 10.0 & 0.0 & 0.01 & 0.26 & 0.61 & 0.78 & 0.01 & 0.26 & 0.62 & 0.80 & 0.04 & 0.24 & 0.26 & 0.81 & 0.01 & 0.25 & 0.60 & 0.79 & 0.04 & 0.19 & 0.26 & 0.80 & 0.04 & 0.20 & 0.26 & 0.80 & 0.04 & 0.45 & 0.80 & 0.92 \\ 
\hline
 & 20 & 1 & 1 & 1 & 10 & 15 & 20 & 5.0 & 10.0 & 0.9 & 0.02 & 0.30 & 0.01 & 0.85 & 0.02 & 0.30 & 0.01 & 0.85 & 0.03 & 0.16 & 0.01 & 0.38 & 0.01 & 0.24 & 0.01 & 0.81 & 0.03 & 0.11 & 0.01 & 0.20 & 0.03 & 0.10 & 0.01 & 0.14 & 0.05 & 0.48 & 0.05 & 0.94 \\ 
 & 20 & 1 & 1 & 1 & 10 & 15 & 20 & 5.0 & 10.0 & 0.7 & 0.02 & 0.30 & 0.02 & 0.83 & 0.02 & 0.31 & 0.02 & 0.84 & 0.04 & 0.17 & 0.03 & 0.41 & 0.02 & 0.26 & 0.01 & 0.81 & 0.04 & 0.14 & 0.03 & 0.25 & 0.04 & 0.14 & 0.03 & 0.24 & 0.06 & 0.47 & 0.06 & 0.92 \\ 
 & 20 & 1 & 1 & 1 & 10 & 15 & 20 & 5.0 & 10.0 & 0.5 & 0.02 & 0.28 & 0.02 & 0.83 & 0.02 & 0.30 & 0.02 & 0.83 & 0.04 & 0.19 & 0.03 & 0.46 & 0.02 & 0.27 & 0.02 & 0.81 & 0.04 & 0.15 & 0.03 & 0.30 & 0.03 & 0.15 & 0.03 & 0.29 & 0.06 & 0.49 & 0.06 & 0.94 \\ 
 & 20 & 1 & 1 & 1 & 10 & 15 & 20 & 5.0 & 10.0 & 0.0 & 0.01 & 0.25 & 0.01 & 0.81 & 0.01 & 0.25 & 0.01 & 0.82 & 0.03 & 0.18 & 0.01 & 0.59 & 0.01 & 0.24 & 0.01 & 0.81 & 0.02 & 0.12 & 0.01 & 0.37 & 0.02 & 0.14 & 0.01 & 0.40 & 0.05 & 0.48 & 0.05 & 0.94 \\ 
\hline
 & 20 & 1 & 4 & 5 & 10 & 13 & 20 & 5.0 & 10.0 & 0.9 & 0.30 & 0.12 & 0.65 & 0.86 & 0.30 & 0.13 & 0.67 & 0.86 & 0.35 & 0.24 & 0.45 & 0.84 & 0.24 & 0.08 & 0.60 & 0.82 & 0.32 & 0.22 & 0.43 & 0.75 & 0.34 & 0.24 & 0.45 & 0.79 & 0.48 & 0.25 & 0.81 & 0.93 \\ 
 & 20 & 1 & 4 & 5 & 10 & 13 & 20 & 5.0 & 10.0 & 0.7 & 0.27 & 0.13 & 0.63 & 0.85 & 0.28 & 0.13 & 0.64 & 0.85 & 0.36 & 0.24 & 0.42 & 0.85 & 0.24 & 0.12 & 0.60 & 0.82 & 0.33 & 0.23 & 0.42 & 0.80 & 0.34 & 0.23 & 0.42 & 0.82 & 0.46 & 0.25 & 0.82 & 0.93 \\ 
 & 20 & 1 & 4 & 5 & 10 & 13 & 20 & 5.0 & 10.0 & 0.5 & 0.26 & 0.11 & 0.63 & 0.83 & 0.27 & 0.12 & 0.65 & 0.83 & 0.32 & 0.21 & 0.43 & 0.85 & 0.24 & 0.10 & 0.61 & 0.81 & 0.31 & 0.20 & 0.43 & 0.80 & 0.31 & 0.21 & 0.43 & 0.82 & 0.45 & 0.24 & 0.83 & 0.93 \\ 
 & 20 & 1 & 4 & 5 & 10 & 13 & 20 & 5.0 & 10.0 & 0.0 & 0.25 & 0.12 & 0.60 & 0.81 & 0.26 & 0.11 & 0.62 & 0.82 & 0.36 & 0.21 & 0.45 & 0.88 & 0.25 & 0.11 & 0.60 & 0.80 & 0.35 & 0.20 & 0.45 & 0.86 & 0.35 & 0.21 & 0.45 & 0.87 & 0.48 & 0.24 & 0.82 & 0.94 \\ 
   \hline
\end{tabular}
}
\end{table}

\subsubsection{Bivariate Dunnett-type test: all-pairs power}
Two versions are considered in this complex situation: i) (\textbf{g}) where all elementary hypotheses must be under $H_1$, ii) (\textbf{p}) where the hypotheses between the endpoints must be under $H_1$, but for any treatment-vs. control comparison, whereby the latter should be more practically relevant.  The simultaneous analysis of co-primary endpoints for comparisons versus control can be performed by a per-treatment IUT (where the inference between the treatment-vs.-control comparisons is performed at Bonferroni level) (pIUT) or a global IUT (i.e. all co-primary endpoints are under $H_1$ for all treatment-vs.-control comparisons) (gIUT). These generic tests are compared with i) modified closure tests on per-treatment level pCTP and global level gCTP (each on OBrien-test CTP) and ii) max-T tests for  per-treatment  (pH) and global level (gH) as well as iii) mmm tests for  per-treatment  (pmmm) and global level (gmmm) (not all simulations shown here). \\

Per definition, all tests are rather conservative, see Table \ref{tab:duall}. The power of all tests decreases with decreasing correlation $\rho$. The global IUT (and the related tests) reveal per definition a rather low power and should be avoided. The partial IUT (pIUT) reveals a higher power than its competitors, however the partial mmm-test (pmmm) provides all individual confidence intervals (or adjusted p-values). 
If only one treatment-vs.-control comparison in only one endpoint is under $H_0$, a very small power results, vice versa.

\begin{table}[H]
\caption{Bivariate Dunnett-type comparisons: all-pairs power}
\centering
\label{tab:duall}
\scalebox{0.5}{
\begin{tabular}{rrrrrrrrrrr|rr|rr|rrrr|rrrr}
  \hline
 & n & ma1 & ma2 & ma3 & mb1 & mb2 & mb3 & sa & sb & rho & \textbf{p}maxT & \textbf{g}maxT & \textbf{p}mmm & \textbf{g}mmm & \textbf{p}IUT  & \textbf{p}CTP & \textbf{g}CTP & \textbf{g}IUT & Y11 & Y12 & Y21 & Y22 \\
  \hline
	
$H_0$ & 20 & 1 & 1 & 1 & 10 & 10 & 10 & 5.0 & 10.0 & 0.9 & 0.01 & 0.00 & 0.02 & 0.00 & 0.03 & 0.02 & 0.01 & 0.01 & 0.05 & 0.05 & 0.05 & 0.05 \\

 & 20 & 1 & 1 & 1 & 10 & 10 & 10 & 5.0 & 10.0 & 0.7 & 0.00 & 0.00 & 0.01 & 0.00 & 0.01 & 0.01 & 0.00 & 0.00 & 0.05 & 0.05 & 0.05 & 0.05 \\
 & 20 & 1 & 1 & 1 & 10 & 10 & 10 & 5.0 & 10.0 & 0.5 & 0.00 & 0.00 & 0.00 & 0.00 & 0.01 & 0.00 & 0.00 & 0.00 & 0.05 & 0.04 & 0.05 & 0.05 \\
 & 20 & 1 & 1 & 1 & 10 & 10 & 10 & 5.0 & 10.0 & 0.1 & 0.01 & 0.00 & 0.00 & 0.00 & 0.00 & 0.00 & 0.00 & 0.00 & 0.05 & 0.06 & 0.05 & 0.05 \\
   \hline \hline
$H_{1111}$ & 20 & 1 & 2 & 5 & 10 & 15 & 20 & 5.0 & 10.0 & 0.9 & 0.30 & 0.07 & 0.66 & 0.07 & 0.71 & 0.59 & 0.15 & 0.15 & 0.16 & 0.48 & 0.81 & 0.93 \\
 & 20 & 1 & 2 & 5 & 10 & 15 & 20 & 5.0 & 10.0 & 0.7 & 0.29 & 0.05 & 0.61 & 0.05 & 0.68 & 0.59 & 0.12 & 0.12 & 0.15 & 0.46 & 0.82 & 0.93 \\
 & 20 & 1 & 2 & 5 & 10 & 15 & 20 & 5.0 & 10.0 & 0.5 & 0.28 & 0.03 & 0.57 & 0.04 & 0.65 & 0.56 & 0.12 & 0.12 & 0.17 & 0.46 & 0.79 & 0.92 \\
 & 20 & 1 & 2 & 5 & 10 & 15 & 20 & 5.0 & 10.0 & 0.0 & 0.26 & 0.01 & 0.50 & 0.01 & 0.61 & 0.49 & 0.07 & 0.07 & 0.15 & 0.46 & 0.78 & 0.94 \\
\hline
 $H_{1111}$ & 20 & 1 & 4 & 5 & 10 & 13 & 20 & 5.0 & 10.0 & 0.9 & 0.28 & 0.11 & 0.68 & 0.12 & 0.71 & 0.62 & 0.22 & 0.22 & 0.46 & 0.24 & 0.81 & 0.94 \\
 & 20 & 1 & 4 & 5 & 10 & 13 & 20 & 5.0 & 10.0 & 0.7 & 0.27 & 0.09 & 0.60 & 0.10 & 0.68 & 0.58 & 0.21 & 0.21 & 0.48 & 0.27 & 0.79 & 0.92 \\
 & 20 & 1 & 4 & 5 & 10 & 13 & 20 & 5.0 & 10.0 & 0.5 & 0.28 & 0.06 & 0.54 & 0.06 & 0.64 & 0.54 & 0.16 & 0.16 & 0.47 & 0.24 & 0.78 & 0.93 \\
 & 20 & 1 & 4 & 5 & 10 & 13 & 20 & 5.0 & 10.0 & 0.0 & 0.29 & 0.01 & 0.52 & 0.02 & 0.65 & 0.53 & 0.10 & 0.10 & 0.45 & 0.24 & 0.83 & 0.94 \\
\hline

 $H_{1111}$ & 20 & 1 & 4 & 5 & 10 & 15 & 20 & 5.0 & 10.0 & 0.9 & 0.38 & 0.27 & 0.61 & 0.28 & 0.66 & 0.53 & 0.43 & 0.44 & 0.52 & 0.50 & 0.76 & 0.92 \\
 & 20 & 1 & 4 & 5 & 10 & 15 & 20 & 5.0 & 10.0 & 0.7 & 0.35 & 0.13 & 0.61 & 0.14 & 0.68 & 0.58 & 0.30 & 0.31 & 0.45 & 0.46 & 0.79 & 0.92 \\
 & 20 & 1 & 4 & 5 & 10 & 15 & 20 & 5.0 & 10.0 & 0.5 & 0.36 & 0.10 & 0.55 & 0.10 & 0.65 & 0.54 & 0.25 & 0.26 & 0.46 & 0.46 & 0.78 & 0.92 \\
 & 20 & 1 & 4 & 5 & 10 & 15 & 20 & 5.0 & 10.0 & 0.0 & 0.38 & 0.05 & 0.53 & 0.05 & 0.65 & 0.53 & 0.19 & 0.20 & 0.47 & 0.43 & 0.81 & 0.92 \\
\hline \hline
 $H_{0011}$& 20 & 1 & 1 & 5 & 10 & 15 & 20 & 5.0 & 10.0 & 0.9 & 0.31 & 0.02 & 0.68 & 0.02 & 0.71 & 0.61 & 0.05 & 0.05 & 0.05 & 0.49 & 0.82 & 0.93 \\
 & 20 & 1 & 1 & 5 & 10 & 15 & 20 & 5.0 & 10.0 & 0.7 & 0.26 & 0.02 & 0.61 & 0.02 & 0.68 & 0.58 & 0.06 & 0.06 & 0.06 & 0.47 & 0.79 & 0.94 \\
 & 20 & 1 & 1 & 5 & 10 & 15 & 20 & 5.0 & 10.0 & 0.5 & 0.26 & 0.01 & 0.58 & 0.01 & 0.67 & 0.56 & 0.05 & 0.05 & 0.05 & 0.48 & 0.83 & 0.93 \\
 & 20 & 1 & 1 & 5 & 10 & 15 & 20 & 5.0 & 10.0 & 0.0 & 0.24 & 0.00 & 0.48 & 0.00 & 0.61 & 0.49 & 0.02 & 0.02 & 0.05 & 0.48 & 0.80 & 0.94 \\
\hline
$H_{0101}$ & 20 & 1 & 1 & 1 & 10 & 15 & 20 & 5.0 & 10.0 & 0.9 & 0.32 & 0.00 & 0.04 & 0.00 & 0.05 & 0.02 & 0.01 & 0.01 & 0.05 & 0.49 & 0.04 & 0.91 \\
 & 20 & 1 & 1 & 1 & 10 & 15 & 20 & 5.0 & 10.0 & 0.7 & 0.26 & 0.00 & 0.03 & 0.00 & 0.04 & 0.02 & 0.01 & 0.01 & 0.04 & 0.45 & 0.05 & 0.93 \\
 & 20 & 1 & 1 & 1 & 10 & 15 & 20 & 5.0 & 10.0 & 0.5 & 0.24 & 0.00 & 0.02 & 0.00 & 0.04 & 0.02 & 0.01 & 0.01 & 0.04 & 0.46 & 0.05 & 0.94 \\
 & 20 & 1 & 1 & 1 & 10 & 15 & 20 & 5.0 & 10.0 & 0.0 & 0.23 & 0.00 & 0.01 & 0.00 & 0.03 & 0.01 & 0.00 & 0.00 & 0.06 & 0.46 & 0.05 & 0.92 \\
\hline
   \hline
\end{tabular}
}
\end{table}

\section{Evaluation of an example using CRAN packages}
\begin{figure}[htbp]
	\centering
		\includegraphics[width=0.3\textwidth]{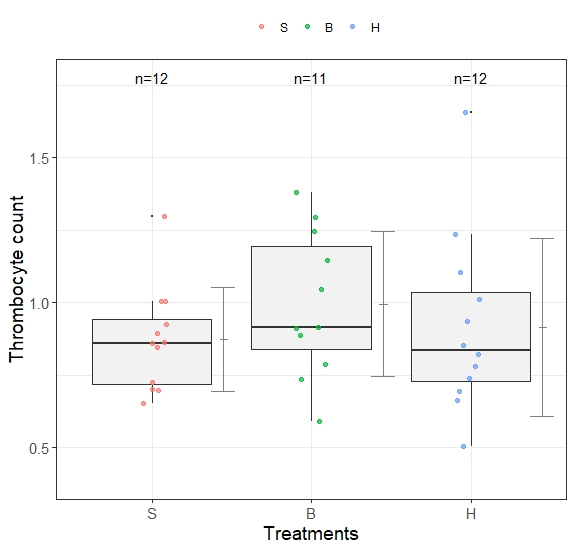}
				\includegraphics[width=0.3\textwidth]{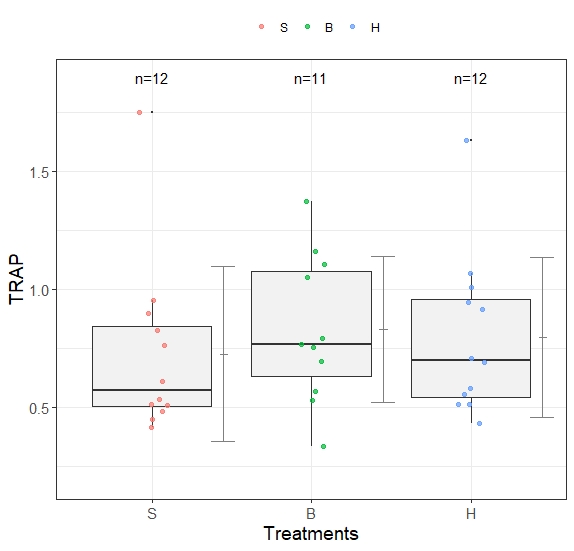}
		\includegraphics[width=0.3\textwidth]{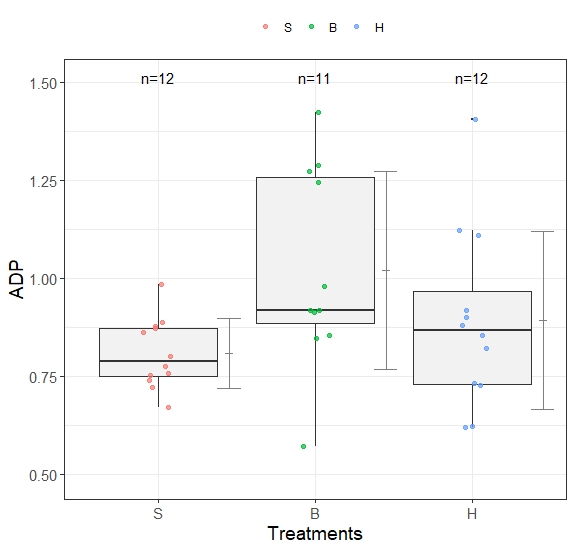}

	\caption{Box plots heart surgery extracorporeal circulation set example}
	\label{fig:bo}
\end{figure}

In heart surgery the standard extracorporeal circulation set (S) is compared with a heparine covering version (H) and a biocompatible surface configuration (B) for the three primary endpoints thrombocyte counts, tartrate resistant acid phosphatase (TRAP) and platelet function (ADP) \cite{Kropf2000}, see the boxplot in Figure \ref{fig:bo}. The raw data are available in the CRAN package SimComp.
Because of the near-to-balanced design, variance heterogeneity is ignored in the following analysis to keep the issue simple. The adjusted p-values for 5 approaches are reported in Table \ref{tab:exa} (bold ... significant). Because within the the CTP's, the global test is not significant, no adjusted p-value can be smaller than 0.05 and therefore the complex tree structure is not shown in this table. 
Only for one primary endpoint, ADP, a significant increase can be shown for the comparison B vs. S using the mmm and MaxT test methods.
(The R-Code is available in the Appendix.)

\begin{table}[H]
\centering\footnotesize
\caption{Adjusted one-sided p-values for 5 approaches} 
\label{tab:exa}
\begin{tabular}{r|l|rr|r|ll}
  \hline
 No & Comparison & mmm & maxT & Bon &CTPps&CTPD \\ 
  \hline
1 & Thromb: B - S & 0.38 & 0.32 & 0.76 & max(0.08,...,0.13)& max(0.06,...,0.13)\\ 
  2 & Thromb: H - S & 0.73 & 0.73& 0.99 &max(0.08,...,0.34)& max(0.06,...,0.34)\\ 
  3 & ADP: B - S & \textbf{0.03} & \textbf{0.04}& 0.05 & max(0.08,...,0.008) & max(0.06,...,0.008)\\ 
  4 & ADP: H - S & 0.44 & 0.38& 0.94 &max(0.08,...,0.16)& max(0.06,...,0.16)\\ 
  5 & TRAP: B - S & 0.58 & 0.59& 0.99 &max(0.08,...,0.23)& max(0.06,...,0.23)\\ 
  6 & TRAP: H - S & 0.69 &0.70 & 0.99 &max(0.08,...,0.31)& max(0.06,...,0.31)\\ 
   \hline
\end{tabular}
\end{table}

\section{Conclusion}
Both CTP and maxT tests can be used to analyze multiple correlated endpoints in both 2 sample and k-sample design for comparing treatments versus control. The expectation of power increase by multiple endpoints (compared to the analysis of single endpoints) is only fulfilled by the concept of any pairs power and only under specific conditions regarding correlations and noncentralities- and only to a modest extent. Considering individual powers, in any configuration a price for multiplicity adjustment have to be paid.
Summarzing the max(max)T-test (mmm) can be recommended: i) it is as powerful as  CTP'susing summarizing global tests, ii) revealing high power when not all endpoint contribute to $H_1$, especially high for not to too low correlations, iii) simultaneous confidence intervals (and adjusted p-value) are available, iv) this approach is available for different scaled endpoints, v) software is available in the CRAN package multcomp, vi) both global and individual claims are available and vii) under particular conditions it represent a competing approach for co-primary endpoints as well.

\footnotesize
\bibliographystyle{plain}

\newpage
\section{Appendix}
\tiny
\begin{verbatim}
library(SimComp)
data(coagulation)
#### max(maxT))
comp2 <- SimTestDiff(data=coagulation, grp="Group", resp=c("Thromb.count","ADP","TRAP"),
                     type="Dunnett", base=3, alternative="greater", covar.equal=FALSE)
summary(comp2)
####### mmm
library(multcomp)
coagulation$Group <- relevel(coagulation$Group, ref = "S")
mod1<-lm(Thromb.count~Group, data=coagulation)
mod2<-lm(ADP~Group, data=coagulation)
mod3<-lm(TRAP~Group, data=coagulation)
TUWI<- glht(mmm(Th=mod1, Ad=mod2, Tr=mod3), mlf(mcp(Group = "Dunnett")), alternative="greater")
summary(TUWI)
################ contrast pairwise
 Cmat2<-c(-1,0,1)  # 3sample pairwise
 Cmat1<-c(-1,1,0)
 PP1<-summary(glht(mod1, linfct = mcp(Group =Cmat1), alternative="greater"))$test$pvalues
 PP2<-summary(glht(mod1, linfct = mcp(Group =Cmat2), alternative="greater"))$test$pvalues
 PP3<-summary(glht(mod2, linfct = mcp(Group =Cmat1), alternative="greater"))$test$pvalues
 PP4<-summary(glht(mod2, linfct = mcp(Group =Cmat2), alternative="greater"))$test$pvalues
 PP5<-summary(glht(mod3, linfct = mcp(Group =Cmat1), alternative="greater"))$test$pvalues
 PP6<-summary(glht(mod3, linfct = mcp(Group =Cmat2), alternative="greater"))$test$pvalues
 PP<-c(PP1,PP2,PP3,PP4,PP5,PP6)
##### OBrien
 library(CombinePValue); library(robustHD); library(simstudy)
 coagulation$S1<-standardize(coagulation$Thromb.count, centerFun = mean, scaleFun = sd)
 coagulation$S2<-standardize(coagulation$ADP, centerFun = mean, scaleFun = sd)
 coagulation$S3<-standardize(coagulation$TRAP, centerFun = mean, scaleFun = sd)
 coagulation$S123<-coagulation$S1+coagulation$S2+coagulation$S3
 mod7 <- lm(S123~Group, data = coagulation)      # OBrien parametric
 PP7<-summary(glht(mod7, linfct = mcp(Group ="Dunnett"), alternative="greater"))$test$pvalues
### distance tests
 XX<-as.matrix(coagulation[, c(2:4)])
 gr<-as.numeric(coagulation$Group)
 grpsel <- c(1,2,3)
 KK <- p_distance_kgroup(XX,gr,grpsel,1,1,1,399) # new Kropf test 1 global 2 01 3 02
 KK4 <- p_distance_kgroup(XX,gr,grpsel,4,1,1,399) # new Kropf test 4 global 2 01 3 02

 ########################
 # Function for distance-based tests with k groups (many-to-one), based on between-statistic
 p_distance_kgroup <- function(x,    # numerical data matrix with n rows(cases)
                               # and p columns (variables),
                               group,       # vector with n rows for group indicators
                               grpsel,      # vector with indices for groups to be compared
                               # (1st: control, others active comparators, supposed to be larger
                               # in one-sided test versions)
                               measure,     # 1: squared Euclidean, 2: Euclidean,
                               # 3: city block, 4: max. absolute component,
                               # 5: Pearson r,
                               sided=2,     # 1: one-sided (group 2 larger under HA), 2: two-sided (default),
                               scalecorr=0, # scale-correction 0 or 1, default without scale correction,
                               nperm=399)   # number of random permutations.
 # The retured vector of p-values combines that from the combined
 # test of all many-to-one comparisons with all pairwise group test.s

 { if (measure==5 & sided==1) {print ('Pearson not available as one-sided version')
   return(NA)}
   nraw <- nrow(x)
   p <- ncol(x)
   k <- length(grpsel)
   ni <- 1:k
   for (ik in 1:k) {ni[ik] <- sum(group==grpsel[ik])};
   n <- sum(ni)
   indexveck <- c(0)
   indexvec <- 1:nraw
   for (ik in 1:k) {indexi <- sort(indexvec*(group==grpsel[ik]))
   indexi <- indexi[(n-ni[ik]+1):n]
   xi=x[indexi,]
   indexveck <- c(indexveck,indexi)}
   indexveck <- indexveck[2:(n+1)]
   x0 <- x[indexveck,]

   betweenmat <- array(data=0, dim=c(n,n,(k-1)))
   nbetween <- 1:(k-1)
   ncum <- ni[1]+1
   for (ik in 2:k)
   {betweenmat[1:ni[1],ncum:(ncum+ni[ik]-1),(ik-1)] <- matrix(data=1,nrow=ni[1],ncol=ni[ik])
   nbetween[(ik-1)] <- sum(betweenmat[,,(ik-1)])
   ncum <- ncum+ni[ik]}


   scalevec <- matrix(data=1, ncol=p, nrow=1)
   if (scalecorr==1) {for (i in 1:p) {scalevec[1,i] <- sqrt(var(x0[,i]))}}
   distmat <- matrix(data=0, nrow=n, ncol=n)
   if (measure==5) {distmat <- cor(t(x0))}
   else
   {for (i in 1:n)
   {for (j in 1:n)
   {diffij <- (x0[i,]-x0[j,])/scalevec
   if (sided==1)   {diffij <- diffij*(diffij>=0)}
   if (measure==1) {aij <- diffij %*% t(diffij)}
   if (measure==2) {aij <- sqrt(diffij %*% t(diffij))}
   if (measure==3) {aij <- sum(abs(diffij))}
   if (measure==4) {aij <- max(abs(diffij))}
   distmat[i,j] <- aij }}}
   permu <- matrix(data=1:n, nrow=1)
   countlargerequal <- 1
   teststatk <- 1:(k-1)
   for (ik in 1:(k-1)) nbetween[ik] <- sum(betweenmat[,,ik])
   if (sided==1 | measure==5) nbetween <- -nbetween

   for (iperm in 0:nperm)
   {distperm <- distmat[permu,permu]
   for (ik in 1:(k-1))
   {teststatk[ik] <- sum(distperm*betweenmat[,,ik])/nbetween[ik]}
   teststat=max(teststatk)
   if (iperm==0) teststat0 <- teststat
   else countlargerequal <- countlargerequal+(teststat>=teststat0)
   permu <- sample(permu)}
   pvalue <- countlargerequal/(nperm+1)

   # here start pairwise tests versus control

   ncum <- 1
   for (ik in 2:k)
   {ncum <- ncum+ni[ik]
   indexi <- c(1:ni[1],ncum:(ncum+ni[ik]-1))
   nni <- length(indexi)
   distmati <- distmat[indexi,indexi]
   betweenmati <- betweenmat[indexi,indexi,(ik-1)]

   permu <- matrix(data=1:nni, nrow=1)
   countlargerequal <- 1
   nbetweeni <- sum(betweenmati)
   if (sided==1 | measure==5) nbetweeni <- -nbetweeni

   for (iperm in 0:nperm)
   {distperm <- distmati[permu,permu]
   teststat <- sum(distperm*betweenmati)/nbetweeni
   if (iperm==0) teststat0 <- teststat
   else countlargerequal <- countlargerequal+(teststat>=teststat0)
   permu <- sample(permu)}

   pvaluei <- countlargerequal/(nperm+1)
   pvalue <- c(pvalue, pvaluei)
   }

   return(pvalue)}


\end{verbatim}

\end{document}